\documentclass[prd,eqsecnum,twocolumn,nofootinbib,superscriptaddress]{revtex4-2}

\usepackage{amsfonts}
\usepackage{mathtools}
\usepackage{amssymb}
\usepackage{amsthm}
\usepackage{bm}
\usepackage{dcolumn}
\usepackage{epsfig}
\usepackage{graphicx}
\usepackage{graphics}
\usepackage[latin1]{inputenc}
\usepackage{latexsym}
\usepackage{rotating}
\usepackage{xcolor}
\usepackage{hyperref} 
\usepackage{calligra} \DeclareMathAlphabet{\mathcalligra}{T1}{calligra}{m}{n} \DeclareFontShape{T1}{calligra}{m}{n}{<->s*[2.2]callig15}{}

\begin{document}
\title{Extreme mass-ratio inspiral of a spinning body into a Kerr black hole\\ I: Evolution along generic trajectories}
\author{Lisa V. Drummond}
\affiliation{Department of Physics and MIT Kavli Institute, MIT, Cambridge, MA 02139 USA}
\author{Alexandra G. Hanselman}
\affiliation{Department of Physics, University of Chicago, Chicago, IL 60637 USA}
\author{Devin R. Becker}
\affiliation{Department of Physics and MIT Kavli Institute, MIT, Cambridge, MA 02139 USA}
\author{Scott A. Hughes}
\affiliation{Department of Physics and MIT Kavli Institute, MIT, Cambridge, MA 02139 USA}

\begin{abstract}
The study of spinning bodies moving in curved spacetime has relevance to binary black hole systems with large mass ratios, as well as being of formal interest.  At zeroth order in a binary's mass ratio, the smaller body moves on a geodesic of the larger body's spacetime.  Post-geodesic corrections describing forces driving the small body's worldline away from geodesics must be incorporated to model the system accurately.  An important post-geodesic effect is the gravitational self-force, which describes the small body's interaction with its own spacetime curvature. This effect includes the backreaction due to gravitational-wave emission that leads to the inspiral of the small body into the black hole. When a spinning body orbits a black hole, its spin couples to spacetime curvature.  This introduces another post-geodesic correction known as the {\it spin-curvature force}.  An osculating geodesic integrator that includes both the backreaction due to gravitational waves and spin-curvature forces can be used to generate a spinning-body inspiral. In this paper, we use an osculating geodesic integrator to combine the leading backreaction of gravitational waves with the spin-curvature force.  Our analysis only includes the leading orbit-averaged dissipative backreaction, and examines the spin-curvature force to leading order in the small body's spin.  This is sufficient to build generic inspirals of spinning bodies, and serves as a foundation for further work examining how to include secondary spin in large-mass-ratio waveform models.
\end{abstract}
\maketitle

\section{Introduction}
\label{sec:intro}
\subsection{Extreme mass-ratio inspirals}
\label{sec:forced motion}

Binary systems with very small mass ratios that inspiral due to the backreaction of gravitational waves (GWs) are called extreme mass-ratio inspirals (EMRIs).  These systems consist of stellar-mass compact objects (of mass $\mu$) orbiting a massive black hole (mass $M$) with mass ratio $\varepsilon \equiv \mu/M$ in the range $10^{-7}\text{--}10^{-4}$.  They produce low-frequency GWs that are expected to be detectable by the planned Laser Interferometer Space Antenna (LISA) \cite{eLISA2013,Barausse2020}.  The detection of GWs from EMRI sources will enable precise measurements of properties of the massive black holes, as well as robustly probe the Kerr nature of the spacetime.  This will be achieved by matching the phase of theoretical waveforms with observed GW data over many thousands to millions of orbits.  Making such measurements will require precise, long-duration waveform models.

The EMRI's small mass ratio $\varepsilon$ means we can treat the binary as the Kerr solution \cite{Kerr1963} plus a perturbation characterizing the smaller body. At zeroth order in $\varepsilon$, the small body's four-momentum  $p^{\alpha}$ obeys the geodesic equation,
\begin{equation}
\frac{Dp^{\alpha}}{d\tau} = 0\;,\label{eq:geodesic}
\end{equation}
where $D/d\tau$ is the covariant derivative along the orbit and $\tau$ is proper time.  Post-geodesic effects, which can be modeled by adding a forcing term to the right-hand side,
\begin{equation}
\frac{Dp^{\alpha}}{d\tau} = f^{\alpha}\;,
\label{eq:forcedgeo}
\end{equation}
describe physics beyond the leading ``free fall'' of a body.  Examples of post-geodesic effects include the self force (arising from the coupling of the small body with its own spacetime curvature) and the spin-curvature force (arising from the coupling of the small body's spin with the background spacetime). These forces are discussed briefly in the following section, and in detail in Sec.\ \ref{sec:forcingterms}.
 
\subsection{Drivers of inspiral evolution}
\label{sec:driversofinspiral}

Schematically, we can write the spacetime metric of a large mass-ratio binary as
\begin{equation}
g_{\mu\nu}^{\text{bin}} = g_{\mu\nu} + \underbrace{h^{(1)}_{\mu\nu}}_\text{ $\mathcal{O(\varepsilon)}$} + \underbrace{h^{(2)}_{\mu\nu}}_\text{ $\mathcal{O}(\varepsilon^2)$} + \mathcal{O}(\varepsilon^3)\;.
\end{equation}
Here, $g_{\mu\nu}^{\text{bin}}$ is the metric describing the binary system, and $g_{\mu\nu}$ is the ``background'' metric, describing its largest member.  The contributions $h^{(1)}_{\mu\nu}$ and $h^{(2)}_{\mu\nu}$ are first- and second-order perturbations arising from the binary's smaller member. If the metric was unperturbed, so that we described the binary's spacetime as just of the background, $g_{\mu\nu}^{\text{bin}} = g_{\mu\nu}$, then the trajectory of a non-spinning body would be given by the geodesic equation in the background $g_{\mu\nu}$ (see Sec. \ref{sec:geodesicsinkerr}).  The trajectory of a \textit{spinning} body in this spacetime would be given by the Mathisson-Papapetrou equations, discussed in detail in Sec.\ \ref{sec:spinningbodymotion}. Including the metric perturbations $h^{(n)}_{\mu\nu}$ introduces self-force effects into the dynamics and leads to the decay of the orbit due to gravitational radiation reaction.

We can write the force on the right-hand side of Eq.\ (\ref{eq:forcedgeo}) as 
\begin{equation}
   f^\alpha = \underbrace{f^{(1)\alpha}}_\text{ $\mathcal{O}(\varepsilon)$} +  \underbrace{f^{(2)\alpha}}_\text{ $\mathcal{O}(\varepsilon^2)$}+\mathcal{O}(\varepsilon^3)\;.
\end{equation}
The first-order term $f^{(1)\alpha}$ arises from $h^{(1)}_{\mu\nu}$ as well as the spin-curvature force, while $f^{(2)\alpha}$ is due to $h^{(2)}_{\mu\nu}$. Explictly, we have
\begin{align}
    f^{(1)\alpha} &= f^{(1)\alpha}_{\text{mono}} + f^\alpha_{\text{SCF}}\;,\\
    f^{(2)\alpha} &= f^{(2)\alpha}_{\text{mono}} + f^\alpha_{\text{dipole}}\;,
\end{align}
where $f^\alpha_{\text{SCF}}$ is the spin-curvature force, and $f^{(1)\alpha}_{\text{mono}}$, $f^{(2)\alpha}_{\text{mono}}$, and $f^{\alpha}_{\text{dipole}}$ are contributions to the self-force. If the small body is a compact object, higher multipoles can generally be neglected, meaning that the pole-dipole approximation is used and the body is described entirely by its mass $\mu$ and spin $S$. The subscript ``mono" denotes that the force is due to the mass of the small body (the body's mass monopole); the subscript ``dipole" denotes that the force is due to the spin of the small body (the body's mass current dipole).  In the detailed analysis we present in this paper, we confine our attention to the first-order self force and to the spin-curvature force.  Before doing so, we provide an overview of different forcing effects, including those which are not in our study.

Many of these terms can also be broken into dissipative and conservative contributions.  For example, the first-order self force can be usefully written as
\begin{equation}
   f^{(1)\alpha}_{\text{mono}} = f_{\text{diss}}^{(1)\alpha} +  f_{\text{cons}}^{(1)\alpha}\;.
\end{equation}
The dissipative contribution, $f_{\text{diss}}^{(1)\alpha}$, includes the leading radiation reaction and secular decay of the orbit; the conservative part of this force, $f_{\text{cons}}^{(1)\alpha}$, describes forcing terms which perturb orbital elements without secular change to the orbit.  Dissipative terms change sign under time reversal; conservative forces are time reversal symmetric.  Although the self force contains both dissipative and conservative pieces, the spin-curvature force as described by the Mathisson-Papapetrou equations is conservative.  For the purposes of the following discussion, it will be useful to divide the other forcing terms similarly, although it must be noted that it is somewhat tricky to split some of the second-order forcing terms into dissipative and conservative pieces:
\begin{align}
   f^{(2)\alpha}_{\text{mono}} &= f_{\text{diss}}^{(2)\alpha} +  f_{\text{cons}}^{(2)\alpha}\;,\\
      f^{\alpha}_{\text{dipole}}  &= f_{\text{dipole,diss}}^{\alpha} +  f_{\text{dipole,cons}}^{\alpha}\;.
\end{align}

In this work, we study very large mass ratio inspirals for which the time scale of orbital evolution is significantly longer than the time scale for individual orbits.  This enables us to use an {\it adiabatic approximation}, which treats the inspiral as an orbit whose elements are secularly decaying due to GW backreaction.  The adiabatic approximation neglects the conservative self force, but provides a framework that allows us to identify post-adiabatic corrections to the leading adiabatic evolution. To build an inspiral in this framework, we break the first-order dissipative self force into an orbit-averaged adiabatic part $f^\alpha_{\text{ad}}$ plus oscillations about this average, $f^\alpha_{\text{oscil}}$:
\begin{align}
    f^{(1)\alpha}_{\text{diss}} = f^\alpha_{\text{ad}} + f^\alpha_{\text{oscil}}\;,
\end{align}
where 
\begin{equation}
f^\alpha_{\text{ad}} = \langle f_{\text{diss}}^{(1)\alpha} \rangle\;,\ \ \ f^\alpha_{\text{oscil}} = f^{(1)\alpha}_{\text{diss}} - \langle f_{\text{diss}}^{(1)\alpha}\rangle\;.
\end{equation}
The angle brackets denote a particular average over the orbit; see Eq.\ (1.4) of Ref.\ \cite{Hughes2021} for a precise definition of this average.  A similar decomposition into orbit-averaged and oscillating pieces can be applied to other forcing terms.  This decomposition introduces a two-time-scale expansion, separating orbit-averaged quantities which evolve slowly, on the system's radiation-reaction timescale, from those which oscillate rapidly, on the system's orbital timescale.  This expansion provides an excellent framework for computing the contributions of rapidly oscillating perturbations to the phase of the gravitational waveform as well as the slowly evolving secular contributions \cite{Hinderer2008}. 

Neglecting the issue of resonances (moments in the inspiral when two of the frequencies are in a small integer ratio, which complicates the averaging needed to define $f_{\text{ad}}$ \cite{Hinderer2008, Flanagan2014, Ruangsri2014, Berry2016}), the influence of each of the post-geodesic forces on the phase of the waveform takes the form
\begin{align}
\Phi &= \underset{\text{adiabatic: $f^{\alpha}_{\text{ad}}$}} {\underbrace{\varphi_0 \varepsilon^{-1}}} + \underset{\substack{\text{post-1-adiabatic:}\; \langle f^\alpha_{\text{SCF}}\rangle + \\ f^{\alpha}_{\text{oscil}} + \langle f^{(1)\alpha}_{\text{cons}} \rangle + \langle f^{(2)\alpha}_{\text{diss}} \rangle + \langle f^{\alpha}_{\text{dipole,diss}} \rangle }} {\underbrace{\varphi_1 \varepsilon^0}}
\nonumber\\
&\qquad + \qquad \ldots\;,
\end{align}
where the $\varphi$ coefficients are dimensionless and do not depend on $\varepsilon$.  The leading-order contribution to inspiral phase arises from the adiabatic, first-order force term $f^{\alpha}_{\text{ad}}$.  Neglecting resonances, the most important sub-leading terms come from post-1-adiabatic order forces, which phase counting analyses have shown must be included in order for the waveform to be accurate enough to match phase with LISA sources (e.g., to serve as ``detection templates'' \cite{Hinderer2008}).  The post-1-adiabatic order contribution to the inspiral phase comes from the oscillatory part of the dissipative first-order force $f^{\alpha}_{\text{oscil}}$, the orbit-averaged conservative part of the first-order self force $\langle f^{(1)\alpha}_{\text{cons}}\rangle$, the orbit-averaged dissipative part of the second-order self force, and the orbit-averaged spin-curvature force.

In this analysis, we only include the influences of the time-averaged dissipative part of the first-order self force, $f^{\alpha}_{\text{ad}}$, and the spin-curvature force, $f^\alpha_{\text{SCF}}$.  This means that we essentially treat the motion using the orbital kinematics of spinning bodies, coupled to the leading gravitational backreaction of point masses.  We neglect all conservative self force effects, the oscillatory dissipative self force, and second-order self force effects.  We also do not include dipole contributions to the self force.  Many examples of the forcing effects we neglect have now been computed.  As we survey in the following section, most of these effects have not yet been implemented in a form that is amenable to large-scale application for studies of inspiral.  The calculations we present below can thus be regarded as the simplest analysis possible of relativistic backreaction on spinning bodies moving on generic Kerr orbits.  This work should serve as a useful point of comparison as implementations of the various other forces mature and can be merged into analyses of this type.


\subsection{Past work}

The first-order, orbit-averaged adiabatic self force, $f^{\alpha}_{\text{ad}} \equiv \langle f^{(1)\alpha}_{\text{diss}}\rangle$, acts to evolve an orbit's energy $E$, axial angular momentum $L_z$, and Carter constant $Q$.  The rates of change of these quantities (often called ``fluxes,'' though strictly speaking the rate of change of $Q$ is not a flux) can be inferred by computing how the orbiting body perturbs the curvature of the binary spacetime.  Most importantly, computing the adiabatic contribution only requires knowledge of the curvature perturbation at null infinity and on the large black hole's event horizon; we do not need these quantities at the orbit itself \cite{Mino2003, Isoyama2019}.  These fluxes, and thus knowledge of how to evolve $(E, L_z, Q)$, can be evaluated along generic orbits around a Kerr black hole to obtain corresponding adiabatic inspirals and waveforms \cite{Drasco2006, Fujita2020, Hughes2021, Chua2021, Katz2021}.

The full first-order self force, including pieces that we neglect in our analysis, have been calculated on bound orbits around a Kerr black hole \cite{vandeMeent2018,Lynch2021} and used to generate non-spinning body inspirals \cite{Warburton2012,Osburn2016,vandeMeent2018_2}.  The spin-curvature force $f^{\alpha}_{\text{SCF}}$ is entirely conservative and given by the Mathisson-Papapetrou-Dixon equations discussed in Sec.\ \ref{sec:MPD}.  It has been proven that the motion of a spinning body under the conservative piece of the self-force is Hamiltonian to first order in mass and spin; the explicit form of this Hamiltonian was also obtained, which will likely be useful for EMRI waveform calculations \cite{Blanco2023}. Inspirals along generic orbits around a non-rotating black hole including both the spin-curvature force as well as the first-order conservative and oscillating dissipative pieces of the self-force were computed in Ref.\ \cite{Warburton2017}. There have been preliminary studies which describe spin-curvature forces using an osculating geodesic formulation for generic orbits; these do not include backreaction due to gravitational radiation \cite{Cunningham2019}. 

For a spinning body orbiting a Kerr black hole, it is also possible to construct the dissipative part of $f^\alpha_{\text{dipole}}$, which enters at post-1-adiabatic order, from the time-averaged energy and angular momentum fluxes computed at infinity and at the black hole horizon \cite{Akcay2020}. The fluxes have been evaluated for circular orbits of spinning bodies in both Schwarzschild \cite{Harms2016_2,Nagar2019} and Kerr spacetimes \cite{Han2010,Harms2016,LukesGerakopoulos2017,Piovano2020}, as well as for eccentric equatorial orbits around a Kerr black hole \cite{Skoupy2021}.  Skoup\'{y} and Lukes-Gerakopoulos used these fluxes to compute the adiabatic inspiral of a spinning body in the equatorial plane of a Kerr black hole \cite{Skoupy2022}.  A study of the effect of a spinning secondary on the self force in a Schwarzschild background with aligned spin and a circular orbit was conducted by Mathews and collaborators \cite{Mathews2022}.  Very recently, Skoup\'{y} and Lukes-Gerakopoulos, with two authors of this paper (LVD and SAH), computed fluxes for generic spinning body orbits for the first time \cite{Skoupy2023}.  

Second-order self-force calculations of $f^{(2)\alpha}_{\text{mono}}$ are just beginning to be applied to astrophysically interesting situations.  Work to date has focused on computing this force in the Schwarzschild spacetime \cite{Pound2020,Warburton2021,Wardell2021}.  This already provides important constraining information which has been exploited to refine the description of binaries in the effective one-body approach \cite{vandemeent2023}.

\subsection{Synopsis of osculating element approach}
\label{sec:osculatingelement}

In this work, we use osculating element integration to solve the forced equations of motion (\ref{eq:forcedgeo}).  This method was first presented in Ref.\ \cite{Pound2008} for osculating Schwarzschild orbits, and was generalized to Kerr in Ref.\ \cite{Gair2011}.  This approach approximates the worldline of a body moving through some spacetime as a sequence of geodesics of that spacetime.  We begin by considering a geodesic that is described by a set of orbital elements $\mathcal{E}^A$ (where the index $A$ ranges from $1$ to $7$, and labels the different elements which we describe in detail below).  We assume that some force drives the true worldline $x^\alpha(\tau)$ away from this geodesic.  We require that this force leaves the motion bounded; this means that we cannot apply this method during the final ``plunge" phase of the inspiral.  At each moment $\tau$, we assume that the worldline under consideration is tangent to a geodesic characterized by elements $\mathcal{E}^A$.  The sequence of geodesics the worldline traverses are called {\it osculating} orbits.  The wordline transitions smoothly from one geodesic to the next, with the corresponding orbital elements $\mathcal{E}^A(\tau)$ adjusting accordingly.  Hence, the elements are dynamical and acquire a dependence on $\tau$.  We thus characterize the forced motion as a geodesic with evolving orbital elements.

One could directly integrate the forced second-order equations given in (\ref{eq:forcedgeo}), rather than reformulating the problem using osculating elements.  There are several reasons why we find it useful to use osculating elements rather than direct second-order integration. First, the orbital elements which we select to parameterize the orbit can typically be chosen such that they provide useful information about the orbital geometry.  For example, as discussed in Sec.\ \ref{sec:paramgeodorbits}, we can uniquely characterize a geodesic in terms of its semi-latus rectum $p$, eccentricity $e$, and inclination angle $I$.  These parameters can serve as the orbital elements used in the osculating element description.  If the force on the right-hand side of Eq.\ (\ref{eq:forcedgeo}) is ``small'' in a useful sense, then changes in the orbital elements can be interpreted very naturally as small modifications to the orbit's geometry.  Because  orbital elements are constant along a geodesic, and only vary in the presence of an additional force, it is also straightforward to distinguish perturbative from non-perturbative effects, as well as to understand them intuitively in terms of changes to the geometry of the orbit.  Although the osculating element formulation does not require that the perturbing force is small, it tends to be particularly useful in this context \cite{Pound2008}. 

Second, the orbital elements naturally separate into two categories, allowing us to easily differentiate conservative and dissipative effects.  The first category (the ``principle orbital elements") includes constants of motion such as energy.  The second category (``positional orbital elements") describe the body's initial position along the geodesic.  Dissipative terms in the perturbing force lead to secular changes in the principle elements; changes in the positional elements arise from conservative terms \cite{Pound2008}.  This framework therefore combines naturally with the two-time-scale method to allow for the construction of an adiabatic inspiral by computing the orbit averaged evolution of the elements.  Physical insight into the effects of the perturbing force can be obscured when we integrate the second-order, forced geodesic equations directly.  The two-time-scale method cannot be implemented as easily as it can be when we use osculating elements.

\subsection{Organization of this paper; conventions and notation}
\label{sec:org}

The remainder of this paper is organized as follows.  Because our osculating orbit scheme is built on Kerr geodesics, we review their properties in Sec.\ \ref{sec:geodesicsinkerr}, describing the general formulation we use in Sec.\ \ref{sec:generalitiesofgeodesics} and our specific parameterization in Sec.\ \ref{sec:paramgeodorbits}.  We discuss the motion of spinning bodies in Sec.\ \ref{sec:spinningbodymotion}, reviewing the basic equations of motion and constants associated with such motion in Secs.\ \ref{sec:MPD} and \ref{sec:spinningbodyconstants}, and then describing in Sec.\ \ref{sec:smallbodyspin} how the analysis simplifies when we consider only the leading order in spin version of this motion.  We conclude this section by discussing parallel transport along Kerr geodesics (Sec.\ \ref{sec:paralleltransport}), briefly describing the tetrad-based framework we adopt from Ref.\ \cite{vandeMeent2019} which describes how the small body's spin evolves along its worldline.

In Sec.\ \ref{sec:oscelementframework} we describe the osculating geodesic framework which underlies our inspiral analysis, describing how to map a worldline to a set of geodesics with evolving elements in Sec.\ \ref{sec:osculatingorbitalelements}), and then laying out the detailed equations we evolve to generate spinning body inspirals in Sec.\ \ref{sec:contraevol}).  These equations take as input forcing terms, which we describe in Sec.\ \ref{sec:forcingterms}, laying out how to incorporate radiation reaction in Sec.\ \ref{sec:RRforcingterms}, and then how to include the spin-curvature coupling in Sec.\ \ref{sec:SCforcingterms}.  Section \ref{sec:numericalsetup} lays out how we numerically construct spinning body inspirals, first presenting the most general framework (Sec.\ \ref{sec:numsetupgen}) before examining how the problem simplifies for the case of motion confined to the equatorial plane (Sec.\ \ref{sec:numsetupeq}).

We present our results in Sec.\ \ref{sec:inspirals}.  We begin by examining several examples of equatorial inspiral in Sec.\ \ref{sec:eqinspirals}, showing how the inspiral of a spinning body follows, on average, the same trajectory in orbital parameter space as a nonspinning body, but that the interaction of its spin with background curvature adds complicating structure to this trajectory.  We show an example of a generic (inclined, eccentric, and with arbitrarily oriented spin) inspiral in Sec.\ \ref{sec:geninspiral}.  The features we saw in equatorial cases appear here as well, though with even more complicating structure thanks to the more complicated nature of generic motion.  We comment that our study of generic inspiral is limited by the paucity of data available describing adiabatic radiation reaction in this limit; though work continues to generate additional such data, we have confined ourselves to the $a = 0.7M$ generic orbit data set that was used in Ref.\ \cite{Hughes2021}.

Throughout this paper, we work in relativist's units with $G = 1 = c$.  A useful conversion factor in these units is $10^6\,M_\odot = 4.926$ seconds $\simeq 5$ seconds.  We use the convention that lowercase Greek indices on vectors and tensors denote spacetime coordinate indices.  Uppercase Latin indices are used on certain quantities to designate elements of a set that holds parameters which describe osculating orbital elements.

\section{Geodesics in Kerr spacetime}
\label{sec:geodesicsinkerr}

\subsection{Generalities}
\label{sec:generalitiesofgeodesics}

In Boyer-Lindquist coordinates, the metric for a Kerr black hole with mass $M$ and spin parameter $a$ is written \cite{Kerr1963,Boyer1967} 
\begin{align}
ds^2 & =-\left(1-\frac{2r}{\Sigma}\right)\,dt^2+\frac{\Sigma}{\Delta}\,dr^2-\frac{4Mar\sin^2\theta}{\Sigma}dt\,d\phi\nonumber\\
 &+\Sigma\,d\theta^2 +\frac{\left(r^2+a^2\right)^2-a^2\Delta\sin^2\theta}{\Sigma}\sin^2\theta\,d\phi^2,\label{eq:kerrmetric}
\end{align}
where 
\begin{equation}
\Delta =r^2-2Mr+a^2\;,\qquad \Sigma =r^2+a^2\cos^2\theta\;.
\end{equation}
This metric has no dependence on coordinates $t$ and $\phi$, and so admits a pair of Killing vectors $\xi_{t}^{\alpha}$ and $\xi_{\phi}^{\alpha}$. A body freely falling in this spacetime therefore has two constants of motion related to these Killing vectors, the energy per unit mass $E$ and axial angular momentum per unit mass $ L_z$:
\begin{align}
E & =-\xi_{t}^{\alpha} u_{\alpha}= -u_{t}\;,\\
L_z & =\xi_{\phi}^{\alpha} u_{\alpha} =  u_{\phi}\;,
\end{align}
where $u^\alpha$ is the 4-velocity of the free falling body. The Kerr metric also possesses a Killing-Yano tensor $\mathcal{F}_{\mu\nu}$ \cite{Penrose1973}, which has the defining property
\begin{equation}
\nabla_{\gamma}\mathcal{F}_{\alpha\beta}+\nabla_{\beta}\mathcal{F}_{\alpha\gamma}=0\;.
\label{eq:KillingYanoDerivs}
\end{equation}
Carter showed that the Killing tensor $K_{\mu\nu}$, defined as the ``square'' of the Killing-Yano tensor via
\begin{equation}
K_{\mu\nu}=\mathcal{F}_{\mu\alpha}{\mathcal{F}_\nu}^{\alpha}\;,
\end{equation}
yields another constant of motion,
\begin{equation}
K = K_{\alpha\beta} u^{\alpha} u^{\beta}\;,
\end{equation}
known as the ``Carter constant'' \cite{Carter1968}.  When $a = 0$, $ K$ is identical to the square of a body's total angular momentum per unit mass.  It is convenient to define a related conserved quantity $Q$, usually also called the Carter constant, by
\begin{align}
  Q &=  K - \left( L_z - a E\right)^2\;.
\label{eq:Qdef}
\end{align}
When $a = 0$, $ Q$ is the square of a body's total angular momentum per unit mass projected into the $\theta = \pi/2$ plane.  The three constants of motion $(E, L_z, Q)$ are one set of ``principle orbital elements" (as discussed in Sec.\ \ref{sec:osculatingelement}) we can use to denote a particular geodesic in the osculating element framework.

The fact that the Kerr spacetime possesses these conserved quantities allows the geodesic equations to be separated as follows \cite{Carter1968}
\begin{align}
\Sigma^2\left(\frac{d{r}}{d\tau}\right)^2 &= [E({r}^2+a^2)-a L_z]^2\nonumber\\
 & \qquad-\Delta[{r}^2+( L_z-a E)^2+Q]\nonumber\\
 & \equiv R({r})\;,\label{eq:geodr}\\
\Sigma^2\left(\frac{d{\theta}}{d\tau}\right)^2 &=  Q-\cot^2{\theta} L_z^2-a^2\cos^2{\theta}(1- E^2)\nonumber\\
 & \equiv\Theta({\theta})\;,\label{eq:geodtheta}\\
\Sigma\frac{d  {\phi}}{d\tau} &= \csc^2{\theta}  L_z + a E\left(\frac{{ r}^2 + a^2}{\Delta} - 1\right) - \frac{a^2 L_z}{\Delta}\nonumber\\
 & \equiv\Phi_r({r})+\Phi_\theta({\theta})\;,\label{eq:geodphi}\\
\Sigma\frac{d  {t}}{d\tau} &=  E\left(\frac{({ r}^2 + a^2)^2}{\Delta} - a^2\sin^2{\theta}\right)\nonumber\\
&\qquad + a  L_z\left(1 - \frac{{ r}^2 + a^2}{\Delta}\right)\nonumber\\
 & \equiv T_r({r})+T_\theta({\theta})\;.\label{eq:geodt}
\end{align}
When the motion is parameterized using proper time $\tau$ as above, equations (\ref{eq:geodr}) -- (\ref{eq:geodt}) do not entirely separate because the quantity $\Sigma(r,\theta)$ couples the radial and polar dynamics. Changing to a time parameter $\lambda$, defined such that $d\lambda = d\tau/\Sigma$, completely decouples the radial and polar motions: 
\begin{align}
\left(\frac{d{ r}}{d\lambda}\right)^2 = R({r})\;,\qquad \left(\frac{d{\theta}}{d\lambda}\right)^2&=\Theta({\theta})\;,
\nonumber\\
\frac{d  {\phi}}{d\lambda} = \Phi_r({r})+\Phi_\theta({\theta})\;,\qquad \frac{d  { t}}{d\lambda}&=T_r({r})+T_\theta({\theta})\;.
\label{eq:geods_mino}
\end{align}
This ``Mino time'' \cite{Mino2003} is thus very useful for describing the orbits of bodies in the vicinity of a Kerr black hole \cite{Mino2003}.  It is straightforward to convert from $\lambda$ to Boyer-Lindquist time $t$, which describes quantities as measured by a distant observer, by using $d t/d\lambda$.

\subsection{Parameterization of geodesic motion}
\label{sec:paramgeodorbits}

Geodesic orbits around a Kerr black hole are contained within a torus of radius $r_2 \le {r} \le r_1$ and polar angle $\theta_- \le {\theta} \le (\pi - \theta_-)$.  We build the bounds on the radial motion into our parameterization by defining
\begin{align}
    {r} & =\frac{p M}{1 + e\cos\chi_r}\;.
    \label{eq:rdef}
\end{align}
We have introduced here $p$, the orbit's semi-latus rectum, and $e$, its eccentricity.  The variable $\chi_r$ is a relativistic version of the Keplerian true anomaly angle that is commonly used to describe orbital dynamics in Newtonian gravity.  We define\footnote{The angle $\chi_r^S$ we use in this analysis is equivalent to $\chi_{r0}$ in Ref.\ \cite{Hughes2021}.  In \cite{Gair2011}, $\psi_0$ is used to denote the initial radial phase, and is equivalent to our $\chi_r^S$, modulo a minus sign.} $\chi_r=\chi_r^F+\chi_r^S$.  The ``$F$'' superscript signifies that $\chi_r^F$ evolves on fast timescales, related to the orbital motion; the ``$S$'' tells us that $\chi_r^S$ evolves on slow timescales, related to the backreaction.  For geodesics (i.e., in the absence of any forcing terms), $\chi_r^S$ is a constant, corresponding to the initial radial phase.  In Section \ref{sec:oscelementframework}, we promote $\chi_r^S$ to a time-varying quantity to account for its slow evolution under the (conservative) spin-curvature force.

As $\chi_r$ varies from 0 to $\pi$, $r$ ranges from periapsis $r_2$ to apoapsis $r_1$, which are defined as
\begin{equation}
    r_1 = \frac{pM}{1 - e}\;,\qquad r_2 = \frac{pM}{1 + e}\;.
\end{equation}
The function $R(r)$ in Eq.\ (\ref{eq:geodr}) is a quartic with four roots ordered such that $r_4 \le r_3 \le r_2 \le { r} \le r_1$. Thus, we can write 
\begin{equation}
R({ r})=(1- E^2)(r_{1}-{ r})({ r}-r_{2})({r}-r_{3})({ r}-r_{4})\;.
\end{equation}
It is convenient to also define $p_3$ and $p_4$ such that
\begin{equation}
    r_3 = \frac{p_3M}{1 - e}\;,\qquad r_4 = \frac{p_4M}{1 + e}\;.
\end{equation}
With this parameterization, we write the radial component of the geodesic equation (\ref{eq:geods_mino}) as a differential equation for $\chi_r$ \cite{Drasco2004}:
\begin{widetext}
\begin{align}
 \frac{d\chi_r}{d\lambda}&=\frac{M\sqrt{1-E^2}\left[(p-p_3)-e(p+p_3\cos\chi_r)\right]^{1/2}\left[(p-p_4)+e(p-p_4\cos\chi_r)\right]^{1/2}}{1-e^2} \nonumber \\ 
 &\equiv X_r^F(\chi_r)\;.
 \label{eq:chir}
\end{align}
\end{widetext}
Remapping the radial dynamics onto the angle $\chi_r$ makes the bounded nature of the motion explicit, allowing for very easy numerical handling of the radial turning points.

Defining ${z} \equiv \cos{\theta}$, we can write $\Theta({\theta})$ from Eq.\ (\ref{eq:geodtheta}) in terms of roots $0 \le z_- \le 1 \le z_+$ \cite{vandeMeent2019}:
\begin{equation}
\Theta({\theta})=\frac{z_1^2-z^2}{1-z^2}\left(z^2_2-a^2(1- E^2)z^2\right)\;.
\end{equation}
This form, taken from Ref.\ \cite{vandeMeent2019}, has the advantage that it allows for straightforward evaluation in the $a\rightarrow 0$ limit.  Turning points of the polar motion occur where ${z} = z_1$, corresponding to when $\theta = \theta_1$ and ${\theta} = \pi - \theta_1$.  The second polar root $z_2$, given by Eq.\ (15) in Ref.\ \cite{vandeMeent2019}, is not actually reached by physical orbits (it generally corresponds to $\cos\theta > 1$, which has no real solution).  We define the inclination angle $I$ as 
\begin{equation}
    I = \pi/2 - \mbox{sgn}(L_z)\theta_1\;.
\end{equation}
We put $x_I \equiv\cos I$, from which we see that $z_1=\sqrt{1-x_I^2}$.  This allows to us to map our polar motion as follows:
\begin{equation}
\cos{\theta} = \sqrt{1 - x_I^2}\cos\chi_\theta = \sin I\cos\chi_\theta\;,
    \label{eq:thdef}
\end{equation}
where $\chi_\theta$ is another relativistic generalization of the ``true anomaly'' angle used in Newtonian orbital dynamics.  As we did for the radial motion, we define\footnote{The angle $\chi_\theta^S$ is equivalent to $\chi_{\theta0}$ used in Ref.\ \cite{Hughes2021}.  In \cite{Gair2011}, $\chi_0$ is used to denote the initial polar phase, and is equivalent to $\chi_\theta^S$ in this analysis, modulo a minus sign.} $\chi_\theta=\chi^F_\theta + \chi_\theta^S$, breaking this anomaly angle into ``fast'' and ``slow'' terms.  In the absence of forcing terms, $\chi_\theta^S$ is a constant, the initial polar phase.  In Section \ref{sec:oscelementframework}, we promote $\chi_\theta^S$ to a time-varying quantity.  Combining the various reparameterizations with the polar geodesic equation (\ref{eq:geodtheta}) yields an equation governing $\chi_\theta$ \cite{vandeMeent2019,Drasco2004}:
\begin{align}
\frac{d\chi_\theta}{d\lambda}&=\sqrt{ z_2^2-a^2(1-E^2)(1 - x_I^2)\cos^2\chi_\theta}\nonumber\\
&\equiv X_\theta^F(\chi_\theta)\;.
\label{eq:chitheta}
\end{align}

We also need expressions for the coordinate time $t$ and axial angle $\phi$ as functions of $\lambda$:
\begin{align}
t(\lambda) &= t_0 + \Gamma \lambda + \Delta t_r [r(\lambda)] + \Delta t_\theta [\theta(\lambda)]\;, \\
\phi(\lambda) &= \phi_0 + \Upsilon^\phi \lambda+\Delta \phi_r [r(\lambda)]+\Delta \phi_\theta [\theta(\lambda)]\;.
\end{align}
The quantities $t_0$ and $\phi_0$ introduced above denote initial conditions.

We define
\begin{align}
\Gamma &= \langle T_r(r) \rangle +\langle T_\theta(\theta) \rangle \label{eq:Gamma}\;,\\
\Upsilon^\phi &= \langle \Phi_r(r) \rangle +\langle \Phi_\theta(\theta) \rangle \label{eq:Upsphi}\;.
\end{align}
The quantity $\Gamma$ is, in an orbit-averaged sense, the rate at which coordinate time $t$ ``ticks'' per unit Mino time $\lambda$; $\Upsilon^\phi$ is a similarly averaged rate at which the axial coordinate advances per unit $\lambda$.  This means that $\Upsilon^\phi$ is the axial orbit frequency conjugate to Mino time $\lambda$.  The averages used in Eqs.\ (\ref{eq:Gamma}) and (\ref{eq:Upsphi}) are defined by Eqs.\ (2.12) and (2.13) in Ref.\ \cite{Hughes2021}.  We also define 
\begin{align}
\Delta t_r [r(\lambda)] &= \int_0^\lambda \left \{T_r[r(\lambda')] - \langle T_r(r)\rangle\right\}d\lambda'\;,
\\
\Delta t_\theta [\theta(\lambda)] &= \int_0^\lambda \left \{T_\theta[\theta(\lambda')] - \langle T_\theta(\theta)\rangle\right\}d\lambda'\;;
\\
\Delta \phi_r [r(\lambda)] &= \int_0^\lambda \left\{\Phi_r[r(\lambda')] - \langle \Phi_r(r)\rangle\right\}d\lambda'\;,
\\
\Delta \phi_\theta [\theta(\lambda)] &= \int_0^\lambda \left\{ \Phi_\theta[\theta(\lambda')] - \langle \Phi_\theta(\theta)\rangle\right\}d\lambda' \;.
\end{align}
We note that Eqs.\ (2.10) and (2.11) of Ref.\ \cite{Hughes2021}, which were intended to be equivalent to the equations above, left out the integrations, incorrectly presenting only the integrands on the right-hand sides of those equations.

Up to initial conditions, a geodesic orbit can be specified by ``principle orbital elements.'' These are either the constants of motion ($E$, $L_z$, $Q$) or the parameters ($p$, $e$, $I$) describing the geometry of the orbit. We can convert between ($E$, $L_z$, $Q$) and ($p$, $e$, $I$) using mappings given in Refs.\ \cite{FujitaHikida2009, vandeMeent2019,Hughes2021}. The initial conditions of the orbit are specified by ``positional orbital elements'' ($\chi_r^S$, $\chi_\theta^S$, $\phi_0$, $t_0$). In order to find the geodesic trajectories for a particular set of orbital elements $\{p, e, I, \chi_r^S, \chi_\theta^S, \phi_0, t_0\}$, we need only solve differential equations for the radial and polar phases $\chi_r$ and $\chi_\theta$, i.e., Eqs.\ (\ref{eq:chir}) and (\ref{eq:chitheta}).

\section{Motion of a spinning body}
\label{sec:spinningbodymotion}

\subsection{Mathisson-Papapetrou-Dixon equations}
\label{sec:MPD}

The motion of a spinning body in curved spacetime obeys the Mathisson-Papapetrou-Dixon (MPD) equations \cite{Papapetrou1951, Mathisson2010,Mathisson2010G_2,Dixon1970}
\begin{align}
\frac{Dp^{\alpha}}{d\tau} & =-\frac{1}{2}{R^\alpha}_{\,\nu\lambda\sigma}u^{\nu}S^{\lambda\sigma}\;,\label{eq:mp1}\\
\frac{DS^{\alpha\beta}}{d\tau} & =p^{\alpha}u^{\beta}-p^{\beta}u^{\alpha}\;.\label{eq:mp2}
\end{align}
Here, the operator $D/d\tau$ denotes a covariant derivative along the worldline that the body follows, the tensor ${R^\alpha}_{\,\nu\lambda\sigma}$ is the Riemann curvature of the spacetime, and $S^{\alpha \beta}$ is a tensor describing the body's spin angular momentum.  The body's 4-momentum is given by
\begin{equation}
p^{\alpha}=\mu u^{\alpha}-u_{\gamma}\frac{DS^{\alpha\gamma}}{d\tau}\;.\label{eq:momvel}
\end{equation}
Notice that a spinning body's 4-momentum is not necessarily parallel to its 4-velocity.

Equations (\ref{eq:mp1}) and (\ref{eq:mp2}) on their own do not constitute a closed system of equations. We impose an additional constraint called the spin supplementary condition (SSC) which closes the system by fixing the internal degrees of freedom associated with the extended structure of the small body. The SSC chooses one of the many worldlines passing through the small body; the choice of wordline from this set is arbitrary. Because there is no natural choice of SSC in general, we use the Tulczyjew SSC \cite{Tulczyjew1959} in this work,
\begin{equation}
p_{\alpha}S^{\alpha\beta}=0\;,\label{eq:TD}
\end{equation} 
since it is commonly-used for GW analysis applications.  As we'll see shortly, this SSC has a particularly simple interpretation when we linearize the MPD equations in the small body's spin.

\subsection{Constants of motion}
\label{sec:spinningbodyconstants}

The spinning body's worldline admits a constant of motion for each spacetime Killing vector $\xi^{\alpha}$, given by
\begin{equation}
\mathcal{C}=p_{\alpha}\xi^{\alpha}-\frac{1}{2}S^{\alpha\beta}\nabla_{\beta}\xi_{\alpha}\;.
\end{equation}
For a spinning body in Kerr, this yields two conserved quantities, energy and axial angular momentum per unit mass:
\begin{align}
E^S & = -u_t+\frac{1}{2\mu}\partial_{\beta}g_{t\alpha}S^{\alpha\beta} \label{eq:Espin},\\ 
L_z^S & = u_{\phi}-\frac{1}{2\mu}\partial_{\beta}g_{\phi\alpha}S^{\alpha\beta}.\label{eq:Lspin}
\end{align}
No Carter-type integral of the motion exists for spinning bodies in general, although an analogue of this constant exists at linear order in the small body's spin \cite{Rudiger1981}.  It has recently been shown that a Carter-like integral exists up to second-order in the small body's spin for a test body possessing exactly the spin-induced quadrupole moment expected for a Kerr black hole  \cite{2023CompereDruart}.

We define a spin vector from the spin tensor by
\begin{equation}
S^{\mu}=-\frac{1}{2\mu}{\epsilon^{\mu\nu}}_{\alpha\beta}p_{\nu}S^{\alpha\beta}\;,
\label{eq:spinvec}
\end{equation}
where
\begin{equation}
\epsilon_{\alpha\beta\gamma\delta}=\sqrt{-g}[\alpha\beta\gamma\delta]\;
\end{equation}
and $[\alpha\beta\gamma\delta]$ is the totally antisymmetric symbol. The magnitude of the spin vector $S$ is defined by
\begin{equation}
S^2=S^{\alpha}S_{\alpha}=\frac{1}{2}S_{\alpha\beta}S^{\alpha\beta}\;\label{eq:smag}
\end{equation}
and is conserved along the spinning body's wordline. 

\subsection{Leading order in small body's spin}
\label{sec:smallbodyspin}

It is useful to express the magnitude of the small body's spin vector $S^2=S^\alpha S_\alpha$ in terms of a dimensionless spin parameter $s$:
\begin{equation}
S=s\mu^2\;.\label{eq:dimesionless}
\end{equation}
The inequality $0 \le s \le 1$ holds\footnote{Another common convention is to write the small body's spin as $S = \sigma \mu M$; the Kerr limit then puts $0 \le \sigma \le \mu/M$.  In this convention, the parameter $\sigma$ is thus of order the mass ratio, which can be helpful for keeping track of the order of different terms in the equations of motion.} if the smaller body is itself a Kerr black hole.  It follows that $S \le \mu^2$, which means that effects at linear order in $S$ are quadratic in the system's mass ratio, and terms at $\mathcal{O}(S^2)$ or higher are negligible in an EMRI context. The linear-in-spin behavior is of astrophysical interest and enters the system's dynamics at the same order as important self force effects \cite{Pound2021}.

Neglecting terms beyond $\mathcal{O}(S)$, the Tulczyjew SSC (\ref{eq:TD}) yields the useful relation
\begin{equation}
p^{\alpha}=\mu u^{\alpha}\;.
\label{eq:momvelfirstorder}
\end{equation}
This amounts to the orbit's 4-velocity and 4-momentum being parallel and $p^\alpha p_\alpha = -\mu^2$ being a constant of motion in this limit.  Combining Eqs.\ (\ref{eq:TD}) and (\ref{eq:spinvec}) with the Tulczyjew SSC tells us that
\begin{equation}
    S^\alpha u_\alpha = 0
\end{equation}
at leading order in spin, a familiar result from special relativistic analyses of spinning bodies.

A generalization of the Carter constant,
\begin{equation}
K^S=K_{\alpha\beta}u^\alpha u^\beta+\delta\mathcal{C}^S\;,
\label{eq:Kspin}
\end{equation}
is conserved at linear order in $S$ \cite{Rudiger1981}, where
\begin{equation}
\delta\mathcal{C}^S= -\frac{2}{\mu^2} p^{\mu}S^{\rho\sigma}\left( {\mathcal{F}^\nu}_{\sigma}\nabla_{\nu}\mathcal{F}_{\mu \rho } - {\mathcal{F}_\mu}^\nu\nabla_{\nu}\mathcal{F}_{\rho\sigma}\right)\;.
\label{eq:Cspin}
\end{equation}

Finally, the Mathisson-Papapetrou equations reduce to
\begin{align}
\frac{Du^{\alpha}}{d\tau} & =-\frac{1}{2\mu}{R^\alpha}_{\,\nu\lambda\sigma}u^{\nu}S^{\lambda\sigma}\;,
\label{eq:mp1linear}\\
\frac{DS^{\alpha\beta}}{d\tau} &= 0\;\label{eq:mp2linear}
\end{align}
at this order.

\subsection{Parallel transport along Kerr geodesics}
\label{sec:paralleltransport}

When we combine Eqs.\ (\ref{eq:mp2linear}) and (\ref{eq:spinvec}), we obtain the equation for spin vector parallel transport along the wordline:
\begin{equation}
\frac{DS^{\mu}}{d\tau}=0\;.
\label{eq:mp2linear2}
\end{equation}
Van de Meent \cite{vandeMeent2019} has provided a closed-form solution for the parallel transport of a vector along a geodesic.  We use this solution to solve Eq.\ (\ref{eq:mp2linear2}), describing the kinematics of our spinning body to leading order in $S$.  As first outlined by Marck \cite{Marck1983, Marck1983_2, Kamran1986}, this approach begins by constructing a set of tetrad legs $\{e_{0\alpha}, e_{1\alpha}, e_{2\alpha}, e_{3\alpha}\}$ which serve as basis objects for our parallel-transported vector.  We begin by defining $e_{0\alpha} \equiv u_\alpha$, taking the first leg of the tetrad to be the tangent to the geodesic itself.  Next, we observe that the vector
\begin{equation}
   \mathcal{L}^\nu=\mathcal{F}^{\mu\nu} u_{\mu}\;,
    \label{eq:orbangmomdef}
\end{equation}
often called the geodesic's orbital angular momentum vector, is also parallel transported along Kerr geodesics.  It is simple to show that the Carter constant $K = \mathcal{L}^\alpha\mathcal{L}_\alpha$; using this to normalize, we define $e_{3\alpha} \equiv\mathcal{L}_\alpha(\lambda)/\sqrt{ K}$ as another leg of the tetrad.

To complete the tetrad, we need to find two vectors orthogonal to $e_{0\alpha}$ and $e_{3\alpha}$.  Equations (50) and (51) of Ref.\ \cite{vandeMeent2019} define two vectors $\tilde{e}_{1\alpha}$ and $\tilde{e}_{2\alpha}$ which satisfy this requirement, but are not parallel transported along the geodesic.  This is rectified by defining \cite{vandeMeent2019}
\begin{align}
    e_{1\alpha}(\lambda) &= \cos\psi_p(\lambda)\,\tilde{e}_{1\alpha}(\lambda) + \sin\psi_p(\lambda)\,\tilde{e}_{2\alpha}(\lambda)
    \label{eq:tetradleg1}\\
    e_{2\alpha}(\lambda) &= -\sin\psi_p(\lambda)\,\tilde{e}_{1\alpha}(\lambda) + \cos\psi_p(\lambda)\,\tilde{e}_{2\alpha}(\lambda)\;,
    \label{eq:tetradleg2}
\end{align}
and requiring that the precession phase $\psi_p(\lambda)$ obeys
\begin{equation}
\frac{d\psi_p}{d\lambda} = \sqrt{K}\left(\frac{(r^2 + a^2)E - a L_z}{K + r^2} + a\frac{L_z - a(1 - z^2)E}{K - a^2z^2}\right)\;.
\label{eq:precphaseeqn}
\end{equation}
The tetrad we find at the end of this procedure has legs $\{e_{0\alpha}(\lambda), e_{1\alpha}(\lambda), e_{2\alpha}(\lambda), e_{3\alpha}(\lambda)\}$ which are parallel transported along Kerr geodesics.  Our spin vector can then be written
\begin{equation}
    S_\alpha = S^0 e_{0\alpha}(\lambda) + S^1 e_{1\alpha}(\lambda) + S^2 e_{2\alpha}(\lambda) + S^3 e_{3\alpha}(\lambda)\;,
    \label{eq:spinvectetrad}
\end{equation}
where $\{S^0, S^1, S^2, S^3\}$ are all constants and $S_\alpha$ satisfies Eq.\ (\ref{eq:mp2linear2}) by construction.

As already noted, the Tulczyjew SSC implies that $S_\alpha u^\alpha = 0$.  With our definition $e_{0\alpha}(\lambda) =  u_\alpha$, we see that we must have $S^0 = 0$.  The definition $e_{3\alpha}(\lambda)=  \mathcal{L}_\alpha(\lambda)/\sqrt{K}$ tells us that $S^3 \equiv s_\parallel$ denotes components of the small body's spin parallel (or antiparallel) to the orbital angular momentum vector $\mathcal{L}_\alpha(\lambda)$. Correspondingly,  $S^1$ and $S^2$ describe components that lie in the orbital plane, perpendicular to the orbital angular momentum. 

\section{Osculating element framework}
\label{sec:oscelementframework}

We now describe forced motion as evolution through a sequence of geodesic orbits, showing how the forcing terms lead to evolution of the orbital elements which characterize geodesics.

\subsection{Evolution of the orbital elements $\mathcal{E}^A$}
\label{sec:osculatingorbitalelements}

The geodesic equation
\begin{equation}
    \frac{d^2x^\alpha}{d\tau^2}= - \Gamma^\alpha_{\beta\gamma}\frac{dx^\beta}{d\tau}\frac{dx^\gamma}{d\tau}\;
\end{equation}
can be written 
\begin{equation}
\label{eq:ddotrgeo}
\ddot x^\alpha = a^\alpha_{\rm geo}\;,
\end{equation}
where overdot denotes $d/d\tau$.  As observed in Sec.\ \ref{sec:paramgeodorbits}, bound Kerr geodesics can be described by seven parameters:
\begin{equation}
   \mathcal{E}^A \doteq \{p, e, I, \chi_r^S, \chi_\theta^S, \phi_0, t_0\}\;.
\end{equation}
The capital Latin indices introduced here range from $1$ to $7$; the symbol $\doteq$ here means ``the components on the left-hand side are given by the elements of the set on the right-hand side.''  In this set, $p$, $e$ and $I$ are the principle orbital elements describing the geometry of the orbit and $\chi_r^S$, $\chi_\theta^S$, $\phi_0$, and $t_0$ are the positional orbital elements that specify initial conditions.

The parameters $\mathcal{E}^A$ are strictly constant along a geodesic worldline and can be expressed as functions of spatial position and spatial velocity in an orbit.  In other words, we can write
\begin{equation}
    \mathcal{E}^A = \mathcal{E}^A(x^\alpha, \dot x^\alpha)\;.
\end{equation}
We can thus use the chain rule to write
\begin{equation}
\label{eq:dotI}
    \dot{\mathcal{E}}^A = \frac{\partial\mathcal{E}^A}{\partial x^\alpha}\dot x^\alpha + \frac{\partial\mathcal{E}^A}{\partial \dot x^\alpha}\ddot x^\alpha\;.
\end{equation}
Using Eq.\ (\ref{eq:ddotrgeo}) and requiring $\mathcal{E}^A$ to be constant along a geodesic trajectory, we obtain
\begin{equation}
\label{eq:dotIgeo}
    \dot{\mathcal{E}}^A = \frac{\partial\mathcal{E}^A}{\partial x^\alpha}\dot x^\alpha + \frac{\partial\mathcal{E}^A}{\partial \dot x^\alpha}a^\alpha_{\rm geo} = 0\;.
\end{equation}

Consider now forced motion.  In the presence of a perturbing force, the geodesic equation generalizes to
\begin{equation}
    \frac{d^2x^\alpha}{d\tau^2} + \Gamma^\alpha_{\beta\gamma}\frac{dx^\beta}{d\tau}\frac{dx^\gamma}{d\tau} = a^\alpha\;.
    \label{eq:forcedgeod}
\end{equation}
The non-geodesic acceleration $a^\alpha$ is subject to the constraint
\begin{equation}
    a^\alpha u_\alpha = 0\;.
\end{equation}
Equation (\ref{eq:forcedgeod}) can be written
\begin{equation}
\label{eq:forced}
\ddot{x}^\alpha = a^\alpha_{\rm geo} + a^\alpha\;.
\end{equation}

Our aim is to convert Eq.\ (\ref{eq:forced}) into a set of equations for the evolution of orbital elements $\mathcal{E}^A$.  This requires a mapping $\{x^\alpha, \dot x^\alpha \} \rightarrow \mathcal{E}^A$.  We assert that, at each moment along the worldline, a geodesic can be found with the same $(x^\alpha, \dot x^\alpha)$ as the accelerated body.  This assertion is called the {\it osculation condition}; stated plainly, we assert that \cite{Pound2008}
\begin{align}
   x^\alpha(\tau) &= x_{\rm geo}^\alpha(\mathcal{E}^A,\tau)\;,
   \\
   \dot x^\alpha(\tau) &= \dot x_{\rm geo}^\alpha(\mathcal{E}^A,\tau)\;,\label{eq:secondeq}
\end{align}
where $a^\alpha(\tau)$ represents the coordinates of the true worldline, and $x^\alpha_{\rm geo}(\mathcal{E}^A,\tau)$ represents the coordinates of a geodesic worldline with orbital elements $\mathcal{E}^A$.  Note that the time derivative in Eq.\ (\ref{eq:secondeq}) holds $\mathcal{E}^A$ fixed.  Note also that the osculation condition involves 4 components of $x^\alpha$ and 4 components of $\dot x^\alpha$, one of which is constrained either by the condition $a^\alpha u_\alpha = 0$ or $u^\alpha u_\alpha = -1$.  The 8 components plus 1 constraint thus map to the 7 parameters $\mathcal{E}^A$, so the number of orbital elements matches the number of degrees of freedom \cite{Pound2008}.  

Under the influence of a perturbing force which accelerates the worldline relative to a geodesic by $a^\alpha$, the parameters $\mathcal{E}^A$ will not remain constant.  We promote them to dynamical variables called \textit{osculating orbital elements}.  The accelerated trajectory $x^\alpha$ is then described by a sequence of geodesics with parameters
\begin{equation}
    \mathcal{E}^A(t^{\rm i}) \doteq \{p(t^{\rm i}), e(t^{\rm i}), I(t^{\rm i}), \chi_r^S(t^{\rm i}), \chi_\theta^S(t^{\rm i}), \phi_0(t^{\rm i}), t_0(t^{\rm i})\}\;.
\end{equation}
We have introduced here $t^{\rm i}$, Boyer-Lindquist coordinate time $t$ along the inspiral.  We will use this as our parameter along the inspiral worldline.  Other parameter choices could be used (e.g., proper time $\tau$ along the inspiral worldline, or Mino time $\lambda$).  The choice of $t$ is particularly convenient for us, as it is the time measured by distant observers.  (Note that we have written both $\phi_0$ and $t_0$ as though they are promoted to dynamical quantities; we will soon show that the equations governing them do not need to be evolved, and they can be left as constants.)

What remains is to prescribe how to dynamically evolve these elements.  We again use the chain rule and Eq.\ (\ref{eq:forced}) to evaluate $\dot{\mathcal{E}}^A(\tau)$, yielding 
\begin{equation}
\label{eq:dotI2}
    \dot{\mathcal{E}}^A = \frac{\partial\mathcal{E}^A}{\partial x^\alpha}\dot x^\alpha + \frac{\partial\mathcal{E}^A}{\partial \dot x^\alpha}a^\alpha_{\rm geo} + \frac{\partial\mathcal{E}^A}{\partial \dot x^\alpha}a^\alpha\;.
\end{equation}
Taking advantage of Eq.\ (\ref{eq:dotIgeo}), we obtain
\begin{equation}
    \dot{\mathcal{E}}^A = \frac{\partial\mathcal{E}^A}{\partial \dot x^\alpha}a^\alpha\;.
\label{eq:oscelementevolforced}
\end{equation}

Multiplying both sides of Eq.\ (\ref{eq:dotIgeo}) by $\partial x_{\rm geo}^\beta/\partial\mathcal{E}^A$ and both sides of Eq.\ (\ref{eq:oscelementevolforced}) by $\partial \dot x_{\rm geo}^\beta/\partial\mathcal{E}^A$ yields a particularly useful form of these equations:
\begin{align}
\frac{\partial x^\beta_{\rm geo}}{\partial\mathcal{E}^A} \dot{\mathcal{E}}^A &= 0\;,
\label{eq:oscelementevol1} \\
\frac{\partial \dot x^\beta_{\rm geo}}{\partial\mathcal{E}^A}\dot{\mathcal{E}}^A &= a^\beta\;.
\label{eq:oscelementevol2}
\end{align}
To derive Eq.\ (\ref{eq:oscelementevol2}), note that Eq.\ (\ref{eq:secondeq}) implies
\begin{equation}
    \frac{\partial\dot x_{\rm geo}^\beta}{\partial\mathcal{E}^A}\frac{\partial\mathcal{E}^A}{\partial\dot x^\alpha} = {\delta^\beta}_\alpha\;.
\end{equation}
These expressions can be used to derive explicit equations for osculating orbital element evolution, and can be written in either contravariant or covariant form (see Secs.\ III D 1 and 2  of Ref.\ \cite{Gair2011}).  As in Ref.\ \cite{Pound2008}, we use the contravariant formulation, which we outline in the next section.

\subsection{Contravariant evolution equations}
\label{sec:contraevol}

For our analysis, we use the contravariant formulation outlined in Sec.\ III D 2 of Ref.\ \cite{Gair2011} to compute the osculating element evolution.  Expanding Eq.\ (\ref{eq:oscelementevol1}) yields
\begin{align}
    \frac{\partial r}{\partial p}p' + \frac{\partial r}{\partial e}e' + \frac{\partial r}{\partial I}I' + \frac{\partial r}{\partial \chi_r^S}\chi_r^{S\prime} + \frac{\partial r}{\partial \chi_\theta^S} \chi_\theta^{S\prime} &= 0\;,
    \label{eq:rgeodeq}\\
    \frac{\partial \theta}{\partial p}p' + \frac{\partial \theta}{\partial e}e' + \frac{\partial \theta}{\partial I}I' + \frac{\partial \theta}{\partial \chi_r^S}\chi_r^{S\prime} + \frac{\partial \theta}{\partial \chi_\theta^S}\chi_\theta^{S\prime} &= 0\;,
    \label{eq:thetageodeq}\\
    \frac{\partial \phi}{\partial p}p' + \frac{\partial \phi}{\partial e}e' + \frac{\partial \phi}{\partial I}I' + \frac{\partial \phi}{\partial \chi_r^S}\chi_r^{S\prime} + \frac{\partial \phi}{\partial \chi_\theta^S}\chi_\theta^{S\prime} + \phi_0' &= 0\;,
    \label{eq:phigeodeq}\\
    \frac{\partial t}{\partial p}p' + \frac{\partial t}{\partial e}e' + \frac{\partial t}{\partial I}I' + \frac{\partial t}{\partial \chi_r^S}\chi_r^{S\prime} + \frac{\partial t}{\partial \chi_\theta^S}\chi_\theta^{S\prime} + t_0'&=0\;.\label{eq:tgeodeq}
\end{align}
Here the prime represents differentiation with respect to the variable used to parameterize the trajectory; as introduced in the previous section, we use coordinate time along the inspiral, $t^{\rm i}$.


Equations (\ref{eq:phigeodeq}) and (\ref{eq:tgeodeq}), which govern the evolution of the axial offset $\phi_0$ and time offset $t_0$, contain elliptic integrals which are introduced due to terms like $\partial t/\partial p$.  Computing such integrals at each time step introduces additional computational expense.  Instead of evolving Eqs.\ (\ref{eq:phigeodeq}) and (\ref{eq:tgeodeq}), we can find $\phi$ and $t$ along the worldline by using the geodesic expressions computed along the instantaneous orbit, as was done in Refs.\ \cite{Gair2011} and \cite{Pound2008}.  Rewriting Eqs.\ (\ref{eq:geodphi}) and (\ref{eq:geodt})), these equations are
\begin{align}
\frac{d\phi}{d\lambda}&=\Phi_r(r, E, L_z,Q )+\Phi_\theta(\theta, E, L_z, Q) \nonumber \\ 
&=\Phi_r[p(\lambda),e(\lambda),I(\lambda),\chi_r^S(\lambda)]\nonumber \\
&+\Phi_\theta[p(\lambda),e(\lambda),I(\lambda),\chi_\theta^S(\lambda)]\;, \label{eq:phiOGeqn}\\ 
\frac{d t}{d\lambda}&=T_r(r, E, L_z, Q)+T_\theta(\theta, E, L_z, Q) \nonumber\\
&=T_r[p(\lambda),e(\lambda),I(\lambda),\chi_r^S(\lambda)]  \nonumber 
\\&+T_\theta[p(\lambda),e(\lambda),I(\lambda),\chi_\theta^S(\lambda)]\;. \label{eq:tOGeqn}
\end{align}
Integrating up Eqs.\ (\ref{eq:phiOGeqn}) and (\ref{eq:tOGeqn}) for $\phi$ and $t$ along the inspiral is equivalent to solving (\ref{eq:phigeodeq}) and (\ref{eq:tgeodeq}).  Observe that Eqs.\ (\ref{eq:rgeodeq}) -- (\ref{eq:tgeodeq}) arise from Eq.\ (\ref{eq:oscelementevol1}), which in turn arises from (\ref{eq:dotIgeo}). Equation (\ref{eq:dotIgeo}) simply states that the geodesic equation $\ddot x^\alpha = a^\alpha_{\rm geo}$ holds when the osculating elements $\mathcal{E}^A$ are constant.  When $\{p, e, I, \chi_r^S, \chi_\theta^S\}$ are all constant, Eqs.\ (\ref{eq:phiOGeqn}) and (\ref{eq:tOGeqn}) yield geodesic solutions; when $\{p, e, I, \chi_r^S, \chi_\theta^S\}$ are evolving, we obtain the solution for forced motion.

We therefore need only consider Eqs.\ (\ref{eq:rgeodeq}) and (\ref{eq:thetageodeq}).  We rearrange these equations to obtain
\begin{align}
\chi_r^{S\prime} &= \frac{1}{\partial r/\partial\chi_r^S}\left(\frac{\partial r}{\partial p}p' + \frac{\partial r}{\partial e}e' + \frac{\partial r}{\partial I}I'\right) \equiv X_r^S(\mathcal{E}^A)\;,
\label{eq:psi0dash}\\
\chi_\theta^{S\prime} &= \frac{1}{\partial \theta/\partial\chi_\theta^S}\left(\frac{\partial \theta}{\partial p}p' + \frac{\partial \theta}{\partial e}e' + \frac{\partial \theta}{\partial I}I'\right) \equiv X_\theta^S(\mathcal{E}^A)\;.
\label{eq:chi0dash}
\end{align}

We next expand Eq.\ (\ref{eq:oscelementevol2}) just as we expanded (\ref{eq:oscelementevol1}):
\begin{align}
    \frac{\partial \dot r}{\partial p}p'+ \frac{\partial \dot  r}{\partial e}e'+ \frac{\partial \dot  r}{\partial I}I'+ \frac{\partial \dot  r}{\partial \chi_r^S}\chi_r^{S\prime}+ \frac{\partial \dot  r}{\partial \chi_\theta^S}\chi_\theta^{S\prime}&=a^r\tau'\;,
    \label{eq:rdoteq}\\
    \frac{\partial \dot  \theta}{\partial p}p'+ \frac{\partial \dot  \theta}{\partial e}e'+ \frac{\partial \dot \theta}{\partial I}I'+ \frac{\partial \dot \theta}{\partial \chi_r^S}\chi_r^{S\prime}+ \frac{\partial \dot \theta}{\partial \chi_\theta^S}\chi_\theta^{S\prime}&=a^\theta\tau'\;,
    \label{eq:thetadoteq}\\
    \frac{\partial \dot  \phi}{\partial p}p'+ \frac{\partial \dot  \phi}{\partial e}e'+ \frac{\partial \dot \phi}{\partial I}I'+ \frac{\partial \dot \phi}{\partial \chi_r^S}\chi_r^{S\prime}+ \frac{\partial \dot \phi}{\partial \chi_\theta^S}\chi_\theta^{S\prime}&=a^\phi\tau'\;,
    \label{eq:phidoteq}\\
     \frac{\partial \dot  t}{\partial p}p'+ \frac{\partial \dot  t}{\partial e}e'+ \frac{\partial \dot t}{\partial I}I'+ \frac{\partial \dot t}{\partial \chi_r^S}\chi_r^{S\prime}+ \frac{\partial \dot t}{\partial \chi_\theta^S}\chi_\theta^{S\prime}&=a^t\tau'\;, \label{eq:tdoteq}
\end{align}
Following Refs.\ \cite{Pound2008} and \cite{Gair2011}, we use the condition $a^\alpha u_\alpha=0$ to eliminate Eq.\ (\ref{eq:tdoteq}).  Following \cite{Gair2011}, we define the useful expression
\begin{equation}
    \mathcal{L}_b(x) \equiv \frac{\partial\dot x}{\partial b} - \frac{\partial r/\partial b}{\partial r/\partial \chi_r^S}\frac{\partial\dot x}{\partial \chi_r^S} - \frac{\partial \theta/\partial b}{\partial \theta/\partial \chi_\theta^S}\frac{\partial\dot x}{\partial \chi_\theta^S}\;,
\end{equation}
where $b$ denotes $p$, $e$ or $I$ and where $x$ denotes $r$, $\theta$ or $\phi$.  This definition allows us to write Eqs.\ (\ref{eq:rdoteq}) -- (\ref{eq:phidoteq}) in the convenient form
\begin{widetext}
\begin{align}
p'& = \frac{\tau'}{D}\left((\mathcal{L}_e(\theta)\mathcal{L}_I(\phi) - \mathcal{L}_e(\phi)\mathcal{L}_I(\theta))a^r + (\mathcal{L}_I(r)\mathcal{L}_e(r) - \mathcal{L}_I(\phi)\mathcal{L}_e(r))a^\theta + (\mathcal{L}_e(r)\mathcal{L}_I(\theta) - \mathcal{L}_e(\theta)\mathcal{L}_I(r))a^\phi\right)\;,
\label{eq:pdash}\\
e'& = \frac{\tau'}{D}\left((\mathcal{L}_I(\theta)\mathcal{L}_p(\phi) - \mathcal{L}_I(\phi)\mathcal{L}_p(\theta))a^r + (\mathcal{L}_p(r)\mathcal{L}_I(r) - \mathcal{L}_p(\phi)\mathcal{L}_I(r))a^\theta + (\mathcal{L}_I(r)\mathcal{L}_p(\theta) - \mathcal{L}_I(\theta)\mathcal{L}_p(r))a^\phi\right)\;,
\label{eq:edash}\\
I'& = \frac{\tau'}{D}\left((\mathcal{L}_p(\theta)\mathcal{L}_e(\phi) - \mathcal{L}_p(\phi)\mathcal{L}_e(\theta))a^r + (\mathcal{L}_e(r)\mathcal{L}_p(r) - \mathcal{L}_e(\phi)\mathcal{L}_p(r))a^\theta + (\mathcal{L}_p(r)\mathcal{L}_e(\theta) - \mathcal{L}_p(\theta)\mathcal{L}_e(r))a^\phi\right)\;,
\label{eq:Idash}\\
D& = \mathcal{L}_p(r)\left(\mathcal{L}_e(\theta)\mathcal{L}_I(\phi) - \mathcal{L}_I(\theta)\mathcal{L}_e(\phi)\right) - \mathcal{L}_e(r)\left(\mathcal{L}_p(\theta)\mathcal{L}_I(\phi) - \mathcal{L}_I(\theta)\mathcal{L}_p(\phi)\right) + \mathcal{L}_I(r)\left(\mathcal{L}_p(\theta)\mathcal{L}_e(\phi) - \mathcal{L}_p(\phi)\mathcal{L}_e(\theta)\right)\;.
\end{align}
\end{widetext}
We then substitute Eqs.\ (\ref{eq:pdash})--(\ref{eq:Idash}) into (\ref{eq:psi0dash})--(\ref{eq:chi0dash}) in order to obtain the evolution of the phase constants $\chi_r^S$ and $\chi_\theta^S$. 

This gives us a closed system of ordinary differential equations for $p$, $e$, $I$, $\chi_r^S$ and $\chi_\theta^S$ given by Eqs.\ (\ref{eq:pdash}) -- (\ref{eq:Idash}), (\ref{eq:chi0dash}) and (\ref{eq:psi0dash}), along with two auxiliary equations for $t$ and $\phi$, Eqs.\ (\ref{eq:phiOGeqn}) and (\ref{eq:tOGeqn}).  Our phase space is given by $\{p, e, I, \chi_r^S, \chi_\theta^S, \phi, t \}$.

\section{Forcing terms}
\label{sec:forcingterms}

The acceleration $a^\mu$ which enters the right-hand side of Eqs.\ (\ref{eq:pdash}) -- (\ref{eq:Idash}) arises in our analysis from two different effects: the self force, which includes radiation reaction; and the spin-curvature force, which is the coupling between curvature and spin discussed in Sec.\ \ref{sec:spinningbodymotion}.  As discussed in Sec.\ \ref{sec:osculatingelement}, the orbital elements fall into two categories, principle orbital elements ($p$, $e$ and $I$) and positional orbital elements ($\chi_r^S$ and $\chi_\theta^S$, after eliminating $\phi_0$ and $t_0$ as discussed in Sec.\ \ref{sec:contraevol}). Secular changes in the principle elements are due to dissipative effects; secular changes in the positional orbital elements are due to conservative effects.

As discussed in the Introduction, for this study we include only the orbit-averaged dissipative first-order terms in the self-force.  These adiabatic terms correspond to radiation-reaction and lead to secular evolution in the principle orbital elements.  The spin-curvature force is purely conservative and drives a secular evolution of the positional elements.  The osculating orbit scheme, when combined with the spin-curvature force, leads to short timescale oscillations in the principle elements.  These oscillations have a magnitude that scales with the mass ratio, and act on the same period as the orbital motions.

\subsection{Radiation reaction}
\label{sec:RRforcingterms}

The stress-energy tensor $T^{\mu\nu}$ for a spinning body in a curved spacetime is  
\begin{align}
\label{eq:Tmunu}
    T^{\mu\nu} &= \frac{1}{\sqrt{-g}}\biggl[\frac{p^{(\mu} u^{v)}}{u^t}\delta^3\left(\mathbf{x} - \mathbf{z}(t^{\rm i})\right)
    \nonumber \\
     &- \nabla_\alpha\left( \frac{S^{\alpha ( \mu}u^{\nu )}}{u^t}\delta^3\left(\mathbf{x} - \mathbf{z}(t^{\rm i})\right)\right)\biggr]\;,
\end{align}
where $\mathbf{z}(t^{\rm i})$ is the body's worldline parameterized by the inspiral coordinate time $t^{\rm i}$.  Notice that the first term in Eq.\ (\ref{eq:Tmunu}) involves the body's momentum and thus depends on its mass.  This monopole term has no information about the body's spatial extent or orientation.  The second term in (\ref{eq:Tmunu}) depends on the spin tensor.  This dipole term thus encodes information about the small body's orientation.

Secondary spin affects $T^{\mu\nu}$ in two ways.  Perhaps most obviously, the dipole term is directly proportional to the spin tensor.  However, secondary spin also enters the monopole term through the impact of the spin-curvature force on the small body's worldline $\mathbf{z}(t^{\rm i})$.  Both effects should be included for a fully self consistent ``pole-dipole'' analysis of motion and backreaction on spinning orbits.

In this work, we use only the monopole term, evaluated at the osculating geodesic at each moment on the inspiral worldline:
\begin{align}\label{eq:TmunuG}
    T_{\rm geo}^{\mu\nu} &= \frac{1}{\sqrt{-g}}\biggl(\frac{\mu u_{\rm geo}^{\mu} u_{\rm geo}^{v}}{u_{\rm geo}^t}\delta^3(\mathbf{x}-\mathbf{z}_{\rm geo}(t^{\rm i}))\biggr)\;.
\end{align}
Our work thus combines the orbital kinematics of spinning bodies with the gravitational-wave backreaction of point particles, neglecting the role of the spinning body's dipolar structure in the dynamics of the gravitational radiation.  We do this primarily because large data sets describing the monopolar term for many geodesic orbits have recently been assembled \cite{Hughes2021, Katz2021, Chua2021}, and it is not difficult for us to repurpose these data for our analysis.  However, it is also the case that the spin corrections should be small at each moment, and that the ``spinning body orbits with point-particle backreaction'' approximation is a useful first approximation to the inspiral of a spinning body.  Work in progress is now developing tools for computing radiation from spinning test bodies, including recent studies of inspiral for aligned equatorial orbits \cite{Skoupy2021}, and a first study of backreaction on generic orbits \cite{Skoupy2023}.  It will be interesting to compare ``spinning body orbits with point-particle backreaction'' to ``spinning body orbits spinning body backreaction'' to assess the importance of the effects that we neglect in this analysis (for example, ascertaining the importance of ``small'' effects which may accumulate over many cycles, with non-negligible net impact).

\subsubsection{Solving the Teukolsky equation}

We infer the impact of the adiabatic self force by computing GWs and the associated rates of change of the orbital integrals $E$, $L_z$, and $Q$ using the Teukolsky equation \cite{Teukolsky1973}.  See Ref.\ \cite{Hughes2021} for a detailed discussion of the methods we use; we provide a brief overview here.

The Teukolsky equation computes perturbations to the Weyl Newman-Penrose curvature scalar $\psi_4$, defined as
\begin{equation}
    \psi_4=-C_{\alpha\beta\gamma\delta}n^\alpha \bar m^\beta
n^\gamma \bar m ^\delta\;,
\end{equation}
where $C_{\alpha\beta\gamma\delta}$ is the Weyl curvature tensor and $n^\alpha$ and $\bar m ^\alpha$ are legs of the Newman-Penrose null tetrad \cite{Newman1962}:
\begin{align}
    n^\mu&=\frac{1}{2\Sigma}(\varpi^2,-\Delta,0,a)\;,\\
     \bar m^\mu&=-\frac{1}{\sqrt{2}\zeta}(i a \sin \theta,0,-1,i \csc \theta)\;;
\end{align}
the factor $\zeta=r-ia\cos\theta$.  Teukolsky derived the equation governing $\psi_4$ \cite{Teukolsky1973},
\begin{equation}
    _{-2}\mathcal{O}\  _{-2}\Psi = 4\pi\Sigma _{-2}\mathcal{T}\;, \label{eq:TEq}
\end{equation}
where $_{-2}\Psi=\zeta^4\psi_4$, $_{-2}\mathcal{O}$ is a second order partial differential operator, and $_{-2}\mathcal{T}$ is a source term.  Note that we have specialized to spin-weight $s = -2$, a particularly convenient choice for studies of gravitational radiation.  The forms for other spin weights, as well as the explicit form of the source term, are given in Ref.\ \cite{Teukolsky1973}.

We solve Eq.\ (\ref{eq:TEq}) in the frequency domain, writing $\psi_4$ in a Fourier and multipolar expansion
\begin{align}
\label{eq:psi4}
    \psi_4 &= \frac{1}{\zeta^4}\int_\infty^\infty d\omega \sum^\infty_{l=2}\sum^\infty_{m=-l} R_{lm}(r;\omega)
    \nonumber\\
    &\quad \times S_{lm}(\varphi;a\omega) e^{i[m\varphi - \omega(t-t_0)]} \;.
\end{align}
The function $S_{lm}(\vartheta,a\omega)$ introduced here is a spheroidal harmonic of spin-weight $-2$; the field $\psi_4$ so computed is evaluated at the event ($t$, $r$, $\vartheta$, $\varphi$).  This decomposition separates Eq.\ (\ref{eq:TEq}) into a pair of ordinary differential equations governing the spheroidal harmonic and governing the radial dependence $R_{lm}(r,\omega)$.  

As discussed at some length in Ref.\ \cite{Hughes2021}, computing adiabatic inspirals requires knowledge only of $\psi_4$ in the limits $r \to \infty$ and $r \to r_+ = M + \sqrt{M^2 - a^2}$.  The GW strain far from the source is related to $\psi_4$ by 
\begin{equation}
\label{eq:psi4toh}
     \psi_4 = \frac{1}{2}\frac{d^2}{dt^2}(h_+ - ih_\times) \ \text{as} \ r\rightarrow \infty\;.
\end{equation}
By exploiting the Teukolsky-Starobinsky identities \cite{Teukolsky1974}, we can also obtain all the information we need about the secondary's impact on the spacetime for our adiabatic analysis by examining $\psi_4$ at the event horizon, $r \to r_+$.

Identities make it possible to compute fluxes at the horizon using $\psi_4$. The solutions to the Teukolsky equation allow us to construct the rates of change of $E$, $L_z$, and $Q$ from GW backreaction and radiation absorbed by the horizon.  A key point for us is that the radial dependence behaves asymptotically as
\begin{align}
    R_{lm}(r,\omega) &\to Z^\infty_{lm\omega}r^3e^{i\omega r_*} \;, \ \ \ \ \ \ \ \ \  \ \ \ \ \ r\to\infty\;,
    \label{eq:Rteukinf}\\
    R_{lm}(r,\omega) &\to Z^{\rm H}_{lm\omega} \Delta e^{-i(\omega-m\Omega_H)r_*} \;, \ \ \ r \to r_+\;.
    \label{eq:RteukH}
\end{align}
Here the frequency $\Omega_H = a/(2Mr_+)$ is the rotation frequency of the horizon, and $r^*$ is the tortoise coordinate 
\begin{align}
r_*(r) &= r + \frac{Mr_+}{\sqrt{M^2-a^2}}\ln \left(\frac{r - r_+}{2M}\right)\nonumber \\
&\quad\,\,\,\,\, -\frac{Mr_-}{\sqrt{M^2-a^2}}\ln \left(\frac{r - r_-}{2M}\right)\;, 
\end{align}
where $r_- = M - \sqrt{M^2 - a^2}$.

For bound black hole orbits, the frequency $\omega$ in Eq.\ (\ref{eq:psi4}) has support only at discrete harmonics:
\begin{equation}
    \omega \to \omega_{mkn} = m\Omega_\phi + k\Omega_\theta + n\Omega_r\;,
\end{equation}
where $\Omega_{x}$ is the frequency associated with a complete cycle of the orbit's motion in coordinate $x$.  Using this, the amplitudes $Z^{\infty,{\rm H}}_{lm\omega}$ can be further decomposed,
\begin{equation}
    Z^{\infty,{\rm H}}_{lm\omega} = \sum_{k = -\infty}^\infty\sum_{n = -\infty}^\infty Z^{\infty,{\rm H}}_{lmkn}\delta(\omega - \omega_{mkn})\;.
    \label{eq:Zlmkn}
\end{equation}
See Ref.\ \cite{Hughes2021} for all the details of this harmonic decomposition.

The coefficients $Z^{\infty,{\rm H}}_{lmkn}$ contain all the information we need to build adiabatic inspirals and waveforms.  From these coefficients, we compute the rates of change $dE/dt$, $dL_z/dt$, $dQ/dt$ for each geodesic in the osculating sequence.  Each of these quantities break into a contribution from the fields at $r \to \infty$ and at $r \to r_+$.  The energy fluxes $dE/dt$ are given by \cite{Teukolsky1974}:
\begin{align}
    \left(\frac{dE}{dt} \right)^\infty &= \sum_{lmkn}\frac{\left|Z^\infty_{lmkn}\right|^2}{4\pi\omega^2_{mkn}}\;,
    \label{eq:dEinfdt}\\
    \left(\frac{dE}{dt} \right)^{\rm H} &= \sum_{lmkn}\frac{\alpha_{lmkn}\left|Z^{\rm H}_{lmkn}\right|^2}{4\pi\omega^2_{mkn}}\;. 
    \label{eq:dEHdt}
\end{align}
The factor $\alpha_{lmkn}$ appears quite a lot when examining quantities which are evaluated on the event horizon, and is given by
\begin{widetext}
\begin{eqnarray}
\alpha_{lmkn} &=& \frac{256(2Mr_+)^5(\omega_{mkn} - m\Omega_{\rm H})[(\omega_{mkn} - m\Omega_{\rm H})^2 + 4\epsilon^2][(\omega_{mkn} - \Omega_{\rm H})^2 + 16\epsilon^2]\omega_{mkn}^3}{|C_{lmkn}|^2}\;.
\label{eq:alphadef}
\end{eqnarray}
The factors $|C_{lmkn}|^2$ and $\epsilon$ are in turn given by
\begin{eqnarray}
|C_{lmkn}|^2 &=& [(\lambda_{lmkn}^2 + 2)^2 + 4am\omega_{mkn} - 4a^2\omega_{mkn}^2](\lambda_{lmkn}^2 + 36am\omega_{mkn} - 36a^2\omega_{mkn}^2)
\nonumber\\
&+& (2\lambda_{lmkn} + 3)(96a^2\omega_{mkn}^2 - 48am\omega_{mkn}) + 144\omega_{mkn}^2(M^2 - a^2)\;,
\label{eq:Clmknsq}\\
\epsilon &=& \frac{\sqrt{M^2 - a^2}}{4Mr_+}\;.
\label{eq:epsilon}
\end{eqnarray}
\end{widetext}
The angular momentum fluxes $dL_z/dt$ are \cite{Teukolsky1974}:
\begin{align}
     \left(\frac{dL_z}{dt} \right)^\infty &= \sum_{lmkn}\frac{m\left|Z^\infty_{lmkn}\right|^2}{4\pi\omega^3_{mkn}}\;,
     \label{eq:dLzinfdt}\\
     \left(\frac{dE}{dt} \right)^{\rm H}&=\sum_{lmkn}\frac{\alpha_{lmkn}m\left|Z^{\rm H}_{lmkn}\right|^2}{4\pi\omega^3_{mkn}}\;.
     \label{eq:dLzHdt}
\end{align}
The Carter constant ``fluxes''\footnote{Strictly speaking, $dQ/dt$ is not a flux since one cannot isolate a contribution of the rate of change of $Q$ by only examining the GWs emitted from a system.  Equations (\ref{eq:dQinfdt}) and (\ref{eq:dQHdt}) instead involve quantities from the radiation field combined with averaged properties of the orbit.  It is common to call $dQ/dt$ a flux nonetheless, since it enters the adiabatic backreaction analysis identically to the true fluxes $dE/dt$ and $dL_z/dt$.} $dQ/dt$ are computed by averaging the dissipative self force on a geodesic \cite{Sago2006}, and are given by
\begin{align}
    \left(\frac{dQ}{dt} \right)^\infty &= \sum_{lmkn}\left|Z^\infty_{lmkn}\right|^2\frac{\mathcal{L}_{mkn}+k\Upsilon_\theta}{2\pi\omega^3_{mkn}}\;,
    \label{eq:dQinfdt}\\
    \left(\frac{dQ}{dt} \right)^{\rm H}&=\sum_{lmkn}\alpha_{lmkn}\left|Z^{\rm H}_{lmkn}\right|^2\frac{\mathcal{L}_{mkn}+k\Upsilon_\theta}{2\pi\omega^3_{mkn}}\;, 
    \label{eq:dQHdt}
\end{align}
where $\Upsilon_\theta$ is the frequency, conjugate to Mino time $\lambda$, characterizing the polar motion, and where
\begin{equation}
    \label{eq:Lmkn}
    \mathcal{L}_{mkn} = m\langle \cot^2\theta \rangle L_z -a^2\omega_{mkn}\langle \cos^2\theta \rangle E\;.
\end{equation}
The expressions $\langle \cot^2 \theta\rangle$ and $\langle \cos^2 \theta\rangle$ in Eq.\ (\ref{eq:Lmkn}) denote $ \cot^2 \theta$ and $ \cos^2 \theta$ averaged over an orbit using
\begin{equation}
    \langle f_\theta(\theta) \rangle=\frac{1}{\Lambda_\theta}\int_0^{\Lambda_\theta} f_\theta[\theta(\lambda)] \ d\lambda\;.
\end{equation}
To evolve from geodesic to geodesic in the osculating sequence, we impose a balance law:
\begin{equation}
   \left( \frac{d\mathcal{C}}{dt}\right)^{\text{orbit}}=- \left( \frac{d\mathcal{C}}{dt}\right)^{\infty}- \left( \frac{d\mathcal{C}}{dt}\right)^{\rm H}
\end{equation}
for each $\mathcal{C} \in (E, L_z, Q)$.  This balance is equivalent to computing the orbit-averaged, leading-order self force.  Using these to evolve from geodesic to geodesic in the osculating sequence builds the adiabatic inspiral.

To build the gravitational waveform associated with this inspiral, we begin by examining Eq.\ (\ref{eq:psi4}) in the limit $r \to \infty$.  Using Eqs.\ (\ref{eq:Rteukinf}) and (\ref{eq:Zlmkn}) in this limit, (\ref{eq:psi4}) yields the following result for $\psi_4$ along the inspiral:
\begin{align}
    \psi_4(t^{\rm i}) &= \frac{1}{r}\sum_{lmkn}Z^\infty_{lmkn}(t^{\rm i})S_{lm}\left[\vartheta, a\omega_{mkn}(t^{\rm i})\right]
    \nonumber\\
    &\qquad \times e^{i\left[m\varphi - \Phi_{mkn}(t^{\rm i})\right]}\;.
\end{align}
Notice that the asymptotic amplitude $Z^\infty_{lmkn}$ and the mode frequency $\omega_{mkn}$ have become functions of the inspiral time $t^{\rm i}$.  We have also replaced the mode frequency times $t$ with the integrated mode phase:
\begin{equation}
    \Phi_{mkn}(t^{\rm i}) = \int_{t_0}^{t^{\rm i}}\omega_{mkn}(t')dt'\;.
\end{equation}
Combining this with Eq.\ (\ref{eq:psi4toh}) allows us to read out the gravitational waveform generated along the inspiral:
\begin{align}
\label{eq:heq}
    h(t^{\rm i}) &\equiv h_+(t^{\rm i}) - ih_\times(t^{\rm i})\nonumber\\
    &= \frac{1}{r} \sum_{lmkn}A_{lmkn}(t^{\rm i}) S_{lm}[\varphi; a\omega_{mkn}(t^{\rm i})] e^{i[m\varphi-\Phi_{mkn}(t^{\rm i})]}\;,
\end{align}
where 
\begin{equation}
\label{eq:Almkn}
    A_{lmkn}(t^{\rm i}) = -\frac{2Z^\infty_{lmkn}(t^{\rm i})}{\omega^2_{mkn}(t^{\rm i})}\;.
\end{equation}
Further discussion and detailed justification of various steps introduced in this section is given in Ref.\ \cite{Hughes2021}.

\subsubsection{Implementation: Grid structure and interpolation}

\begin{figure*}
\centerline{\includegraphics[scale=0.43]{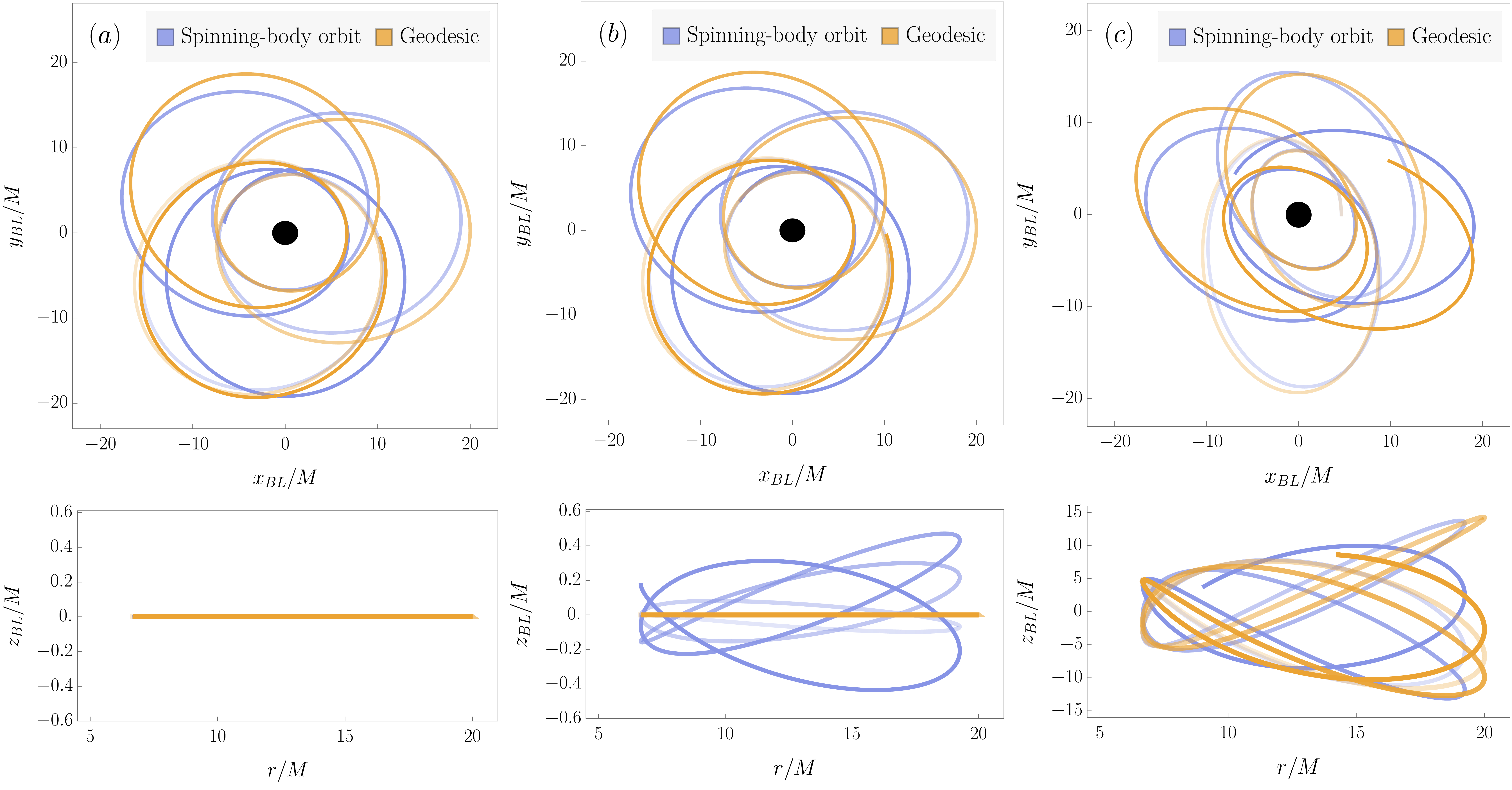}}
    \caption{Comparison of spinning-body (blue) and geodesic (orange) orbit trajectories.  Within each column, the trajectories shown have the same initial conditions.  The top row shows $x_{BL}$-$y_{BL}$ trajectories; the bottom shows trajectories in $r$-$z_{BL}$.  (The coordinates $x_{BL}$ etc.\ are Cartesian-like representations of the Boyer-Lindquist coordinates; see text for precise definitions.)  Increasing opacity of the trajectory curves denotes increasing time. Panel (a) shows equatorial trajectories; for the blue (spinning-body) trajectory, the spin of the small body is aligned with the spin of the larger black hole.  The major difference in the trajectories in this case is the dephasing that occurs because spin-curvature coupling changes the timescales associated with orbital motions.  Panel (b) shows the same geodesic orbit as panel (a) but the spinning-body trajectory corresponds to a small body with its spin misaligned with its orbit.  Notice that the in-plane motion is similar to what we find in panel (a), at least over the time interval shown here, though the motion acquires an out-of-plane motion that is entirely absent from the geodesic case.  Panel (c) shows generic orbits for both cases.  In all panels, the parameters used are $a = 0.7 M$, $p = 10$, $e = 0.5$, $\varepsilon=0.1$, and $s = 1$.  In panels (b) and (c), we put $s_\parallel = 0.9s$ and $\phi_s = \pi/2$; in panel (c), we further put $x_I = 0.6967$.  Here and in many of the other plots, we have used a much less extreme mass ratio than is appropriate for these techniques in order to magnify the effect of spin-curvature coupling physics.} 
    \label{fig:trajectory}
\end{figure*}

We compute the amplitudes $Z^{\infty,{\rm H}}_{lmkn}$ and infer the associated fluxes using GREMLIN, a Teukolsky equation solver implemented in the frequency-domain. This code was developed mainly by author Hughes; see Refs.\ \cite{Hughes2000,Drasco2006} for key details. Some of the methods used in the code have since been updated; see Refs.\ \cite{Fujita2004, Fujita2005, Throwe2010, OSullivan2014, Hughes2021}.

We use datasets consisting of these amplitudes and fluxes that were presented in Ref.\ \cite{Hughes2021}.  All of the data sets we use are evenly spaced in eccentricity, with $\delta e = 0.1$, but are non-uniformly spaced in $p$.  The grid we use is spaced uniformly in $u$, defined as
\begin{equation*}
    u\equiv \frac{1}{\sqrt{p - 0.9p_{\rm LSO}}}\;.
\end{equation*}
Even spacing in $u$ leads to denser coverage in $p$ near $p_{\rm LSO}$, which allows us to account for the rapid variation in certain quantities near the LSO.  We use 40 points between $p_{\rm min} = p_{\rm LSO} + 0.02M$ and $p_{\rm max} = p_{\rm min} + 10M$.  We examine two classes of datasets: equatorial, covering the spin range $a/M \in [0, 0.1, 0.2, \ldots, 0.9, 0.95, 0.99]$ and the eccentricity range $0 \le e \le 0.8$; and one generic dataset, with $a/M = 0.7$, $0 \le e \le 0.4$, and covering the inclination range $-1 \le x_I \le 1$, with $\delta x_I = 2/15$.  See Sec.\ VI of \cite{Hughes2021} for further details about the flux data produced by GREMLIN.  (We comment that additional datasets are under construction, though using a somewhat different grid, more similar to that described in Ref.\ \cite{Chua2021}.)

Our grid data are interpolated across parameter space to provide fluxes as continuous functions of $p$, $e$, and $x_I$ for input to the osculating element scheme.  For the two-dimensional, equatorial dataset, we interpolate directly in Mathematica using a cubic spline interpolation package developed based on numerical methods in \cite{Press2007}; see \cite{Hanselman2020} for more details.  This package was constructed to deal with two-dimensional datasets only.  For the three-dimensional generic dataset, we found it efficient to interpolate first in MATLAB using the \texttt{griddata} function in order to produce a rectangular dataset. We use the \texttt{natural} interpolation method which is a triangulation-based natural neighbor interpolation. This was the highest accuracy method offered by \texttt{griddata} that supported 3D grids; it is designed to be an efficient trade-off between linear and cubic interpolation.  Once we produce the rectangular, more densely-spaced dataset ($\Delta p=0.1$, $\Delta e=0.01$ and $\Delta x_I=0.02$), we then use the \textit{Mathematica} function \texttt{Interpolation} to generate interpolation functions that can be evaluated efficiently as we integrate the osculating geodesic equations.  This interpolation scheme is not intended to be accurate enough to make ``production quality'' waveforms (i.e., which could be used in LISA-related science or data analysis studies), but suffices for the purposes of this analysis.

\subsection{Spin-curvature force}
\label{sec:SCforcingterms}

We augment the adiabatic self force Eq.\ (\ref{eq:mp1linear}), defining the spin-curvature acceleration
\begin{equation}
    a_{\rm SCF}^\alpha \equiv f_{\rm SCF}^\alpha/\mu\equiv -\frac{1}{2\mu}{R^\alpha}_{\,\nu\lambda\sigma}u^{\nu}S^{\lambda\sigma}\;.
\end{equation}
The spin tensor which enters this expression is constructed as discussed in Sec.\ \ref{sec:MPD}, linearizing all quantities in the small body's spin and using the tetrad formulation for parallel transport described in Sec.\ \ref{sec:paralleltransport}.  All terms are evaluated using values corresponding to the instantaneous geodesic in the osculating sequence at each moment along the worldline.

Before turning to our study of inspirals and comparisons of the trajectories followed by spinning and non-spinning bodies, we briefly examine some of the key differences between spinning-body and geodesic orbits; Refs.\ \cite{Witzany2019_2,Drummond2022_2} provide more details.  Spinning-body orbits are qualitatively different from geodesic ones.  If the body's spin is misaligned from the orbit, then its orientation precesses, with a Mino-time frequency $\Upsilon_s$ characterizing this precession; the body's orbital plane likewise precesses at this frequency.  This precession appears in the equations of motion as a variation in the bounds of both the polar and radial libration regions.  Indeed, one finds that the radial and polar motions for a spinning body do not separate when parameterized in Mino time as they do in the geodesic case \cite{Witzany2019_2,Drummond2022_1,Drummond2022_2}.  Finally, a body's spin also shifts the orbital frequencies relative to the orbital frequencies associated with geodesic orbits.  The well-understood frequencies $\Omega_{r,\theta,\phi}$ which characterize geodesic orbits are each shifted by an amount $\propto s_{\parallel}$, the component of the smaller body's spin parallel to its angular momentum.

We first consider equatorial orbits with aligned spin: $s = s_\parallel$, $s_\perp = 0$.  Spinning-body and geodesic orbits are qualitatively the same in this case: motion is constrained to the plane $\theta = \pi/2$, and the radial motion is confined to an interval $r_2 \le r \le r_1$, where $r_2$ and $r_1$ are constants.  Equatorial non-spinning and spinning-body orbits with the same initial conditions are shown in panel (a) of Fig.\ \ref{fig:trajectory}.  Differences in the trajectories do emerge, because the two trajectories have different frequencies associated with both their radial and axial motions.

Qualitative differences become quite noticeable when $s_\perp \neq 0$, so that the small body's spin vector is misaligned.  In this case, the spin vector precesses and the spinning body's orbit oscillates by an amount $\mathcal{O}(S)$ out of the equatorial plane.  For these ``nearly equatorial'' orbits, the radial motion is still constrained to lie between $r_2 \le r \le r_1$, but the polar libration range is modified, with $\theta = \frac{\pi}{2} + \delta \vartheta_S$.  The orbital plane precesses in response to the small body's spin precession, adjusting the turning points of the polar motion depending on the spin precession phase $\psi_p$.  This can be seen in panel (b) of Fig.\ \ref{fig:trajectory}; the orange (non-spinning) worldline is confined to the equatorial plane while the blue (spinning-body) worldline oscillates about the equatorial plane.

Fully generic spinning-body orbits have eccentricity, are inclined with respect to the equatorial plane, and have an arbitrarily oriented small-body spin.  Functions evaluated along generic orbits have structure at harmonics of three frequencies: radial $\Omega_r$, polar $\Omega_{\theta}$, and spin-precessional $\Omega_s$.  We can therefore write functions evaluated along an orbit as a Fourier expansion of the form
\begin{equation}
    f[r,\theta,S^{\mu}] = \sum_{j = -1}^1 \sum_{k,n = -\infty}^\infty f_{jkn} e^{-ij\Omega_st}e^{-in\Omega_rt}e^{-ik\Omega_{\theta}t}\;,
\end{equation}
where $S^{\mu}$ is the small-body's spin vector.  Notice the different index ranges in this sum: only three harmonics of the spin frequency $\Omega_s$ are ever present, while in principle an infinite set of both polar and radial harmonics are present.  (In practice, these sums converge over a finite range, though one must study the system carefully to determine an appropriate truncation point \cite{Drummond2022_2}.)

The coupling of radial, polar and spin-precessional motions for generic spinning-body orbits causes the positions of the radial turning points to depend on $\theta$ and the spin-precession phase $\psi_p$. Similarly, the polar turning points depend on radial position and $\psi_p$, as derived in Ref.\ \cite{Witzany2019_2}.  Panel (c) of Fig.\ \ref{fig:trajectory} shows a generic geodesic (in orange) and spinning-body trajectory (in blue) with the same initial conditions.  The opacity of the curves increases as time advances; this illustrates how the trajectories diverge at late times, as the opacity increases.

\section{Numerical set-up}
\label{sec:numericalsetup}

\subsection{Evolution along generic trajectories}
\label{sec:numsetupgen}

In the generic case, motion is described by the evolution of osculating orbital elements $p$, $e$, $I$, $\chi_r^S$ and $\chi_\theta^S$ which relate to $r$ and $\theta$ as follows:
\begin{align}
    r &= \frac{pM}{1+e \cos(\chi_r^F+\chi_r^S)}\;,
    \\
    \cos{\theta} &= \sin I\cos(\chi_\theta^F+\chi_\theta^S)\;.
\end{align}

In Sec.\ \ref{sec:oscelementframework}, we presented Eqs.\ (\ref{eq:pdash})--(\ref{eq:Idash}) and (\ref{eq:psi0dash})--(\ref{eq:chi0dash}) which describe the evolution of these osculating orbital elements. Remembering that we defined the total radial phase to be $\chi_r=\chi_r^F+\chi_r^S$ and total polar phase to be $\chi_\theta=\chi_\theta^F+\chi_\theta^S$, we must also find $\chi_r^F$ and $\chi_\theta^F$ using the usual geodesic expressions Eqs.\ (\ref{eq:chir}) and (\ref{eq:chitheta}). For completeness and clarity, we repeat these equations explicitly below:
\begin{align}
\chi_r^{F\prime} &= \frac{\lambda'M\sqrt{1-E^2}}{1-e^2}
\nonumber\\
&\times \biggl[\left\{(p - p_3) - e(p + p_3\cos(\chi_r^F + \chi_r^S))\right\}^{1/2}
\nonumber\\
&\times \left\{(p - p_4) + e(p - p_4\cos(\chi_r^F + \chi_r^S))\right\}^{1/2}\biggr]\;, \label{eq:psidash}
\\ 
\chi_\theta^{F\prime}& = \lambda'a(1 - E)\sqrt{z_+ - z_-\cos^2(\chi_\theta^F + \chi_\theta^S)}\;.\label{eq:chidash}
\end{align}
As used elsewhere, the prime in $\lambda'$ in the above equations denotes that Mino-time is differentiated with respect to the parameter used to describe the orbit, which in our analysis will typically be coordinate-time along the inspiral $t^{\rm i}$.  The right-hand side of Eqs.\ (\ref{eq:pdash})--(\ref{eq:Idash}) depends on the perturbing force.  In this analysis, part of the perturbing force is due to spin-curvature coupling; in order to evaluate the spin-curvature force we need to compute the spin precession phase. We find it convenient to simultaneously solve Eq.\ (\ref{eq:precphaseeqn}) for the spin precession phase,
\begin{equation}
\psi_p'=\lambda'\sqrt{K}\left(\frac{(r^2+a^2)E-a L_z}{K+r^2}+a\frac{L_z-a(1-z^2)E}{K-a^2z^2}\right)\;.\label{eq:psipdash}
\end{equation}
Though there exists an analytic closed-form solution to the above differential equation, we find it numerically useful to evolve the spin precession phase explicitly.

Equations (\ref{eq:pdash})--(\ref{eq:Idash}), (\ref{eq:psi0dash})--(\ref{eq:chi0dash}) and (\ref{eq:psidash})--(\ref{eq:psipdash}) comprise the complete set of equations we use to evolve the orbital elements and construct the inspiral. We rewrite these equations more concisely by defining functions $F_p$, $F_e$ and $F_I$ which represent the right-hand side of Eqs.\ (\ref{eq:pdash})--(\ref{eq:Idash}). We divide each of these functions into two pieces: (i) the spin-curvature force piece, with components $F_p^{\textrm{SCF}}$, $F_e^{\textrm{SCF}}$ and $F_I^{\textrm{SCF}}$; and (ii) the radiation-reaction piece arising from the first-order adiabatic self-force, with components $F_p^{\textrm{RR}}$, $F_e^{\textrm{RR}}$ and $F_I^{\textrm{RR}}$. Similarly, we define the functions on the right-hand side of Eqs.\ (\ref{eq:psi0dash}), (\ref{eq:chi0dash}), (\ref{eq:psidash}), (\ref{eq:chidash}) and (\ref{eq:psipdash}) to be $X^S_r$, $X^S_\theta$, $X^F_r$, $X^F_\theta$ and $\Psi_p$ respectively.  Not all of these functions depend on the full set of variables we are solving for; some are functions of only a subset. This motivates us to define three sets of parameters $\vec{\mathcal{A}}(t)$, $\vec{\mathcal{B}}(t)$ and $\vec{\mathcal{C}}(t)$:
\begin{align}
\vec{\mathcal{A}}(t)&=\{p(t),e(t),I(t)\}\\
\vec{\mathcal{B}}(t)&=\{p(t),e(t),I(t),\chi_r^F(t),\chi_r^S(t),\chi_\theta^F(t),\chi_\theta^S(t)\}\\
\vec{\mathcal{C}}(t)&=\{p(t),e(t),I(t),\chi_r^F(t),\chi_r^S(t),\chi_\theta^F(t),\chi_\theta^S(t),\psi_p(t)\}\;.
\end{align}
Rewriting our system of equations in this compact form, we have
\begin{align}
    \frac{dp}{dt}&=F_p^{\textrm{SCF}}[\vec{\mathcal{C}}(t)]+F_p^{\textrm{RR}}[\vec{\mathcal{A}}(t)]\;,\label{eq:dpdteq} \\ 
    \frac{de}{dt}&=F_e^{\textrm{SCF}}[\vec{\mathcal{C}}(t)]+F_e^{\textrm{RR}}[\vec{\mathcal{A}}(t)]\;, \\
    \frac{dI}{dt}&=F_I^{\textrm{SCF}}[\vec{\mathcal{C}}(t)]+F_I^{\textrm{RR}}[\vec{\mathcal{A}}(t)]\;, \\
    \frac{d\chi_r^S}{dt}&=X_r^S[\vec{\mathcal{C}}(t)]\;, \ \ \frac{d\chi_\theta^S}{dt}=X_\theta^S[\vec{\mathcal{C}}(t)]\;, \\
    \frac{d\chi_r^F}{dt}&=X^F_r[\vec{\mathcal{B}}(t)]\;, \ \ \frac{d\chi_\theta^F}{dt}=X^F_\theta[\vec{\mathcal{B}}(t)]\;,\\ \frac{d\psi_p}{dt}&=\Psi_p[\vec{\mathcal{B}}(t)]\;.\label{eq:psipeq}
\end{align}
Note that the functions $X^F_r[\vec{\mathcal{B}}(t)]$, $X^F_\theta[\vec{\mathcal{B}}(t)]$ and $\Psi_p[\vec{\mathcal{B}}(t)]$ are provided in their entirety in this section, and have no explicit dependence on the perturbing force [see Eqs.\ (\ref{eq:psidash})--(\ref{eq:psipdash})].  The functional forms for $F_p$, $F_e$, $F_I$, $X_r^S$ and $X_\theta^S$ are provided in Sec.\ \ref{sec:contraevol} [see Eqs.\ (\ref{eq:pdash})--(\ref{eq:Idash}) and (\ref{eq:psi0dash})--(\ref{eq:chi0dash})]. They depend explicitly on accelerations $a^r$, $a^\theta$ and $a^\phi$; we discussed how we compute components of the perturbing force in the previous section.  We remark that we compared the output of our osculating geodesic integrator with the action-angle version presented in Ref.\ \cite{Lynch2022} for a test case (the gas drag toy model in \cite{Gair2011}) and found excellent agreement.

\subsection{Specializing to equatorial motion}
\label{sec:numsetupeq}

Equatorial spin-aligned inspirals remain confined to the equatorial plane, which reduces the degrees of freedom and thus the number of parameters we need to characterize these systems.  The inclination angle will always be $I = 0^\circ$ (for prograde inspiral) or $I = 180^\circ$ (for retrograde), and there is no need to track the polar phase $\{\chi_\theta^F,\chi_\theta^S\}$.  In this case, the system's motion can be parameterized entirely using orbital elements $\{p,e,\chi_r^F,\chi_r^S\}$:
\begin{align}
    r&=\frac{pM}{1+e \cos(\chi_r^F+\chi_r^S)}\;.
\end{align}
Equations (\ref{eq:dpdteq})--(\ref{eq:psipeq}) become
\begin{align}
    \frac{dp}{dt} &= F_p^{\textrm{SCF}}[\vec{\mathcal{B}}(t)] + F_p^{\textrm{RR}}[\vec{\mathcal{A}}(t)]\;,\label{eq:dpdteq1} \\ 
    \frac{de}{dt} &= F_p^{\textrm{SCF}}[\vec{\mathcal{B}}(t)] + F_e^{\textrm{RR}}[\vec{\mathcal{A}}(t)]\;, \\
    \frac{d\chi_r^S}{dt} &= X_r^S[\vec{\mathcal{B}}(t)]\;, \ \
    \frac{d\chi_r^F}{dt} = X_r^F[\vec{\mathcal{B}}(t)]\;,
\end{align}
where 
\begin{align}
\vec{\mathcal{A}}(t)&=\{p(t),e(t)\}\;,\\
\vec{\mathcal{B}}(t)&=\{p(t),e(t),\chi_r^F(t),\chi_r^S(t)\}\;.
\end{align}

\section{Spinning-body inspirals}
\label{sec:inspirals}
\begin{figure*}
\centerline{\includegraphics[scale=0.53]{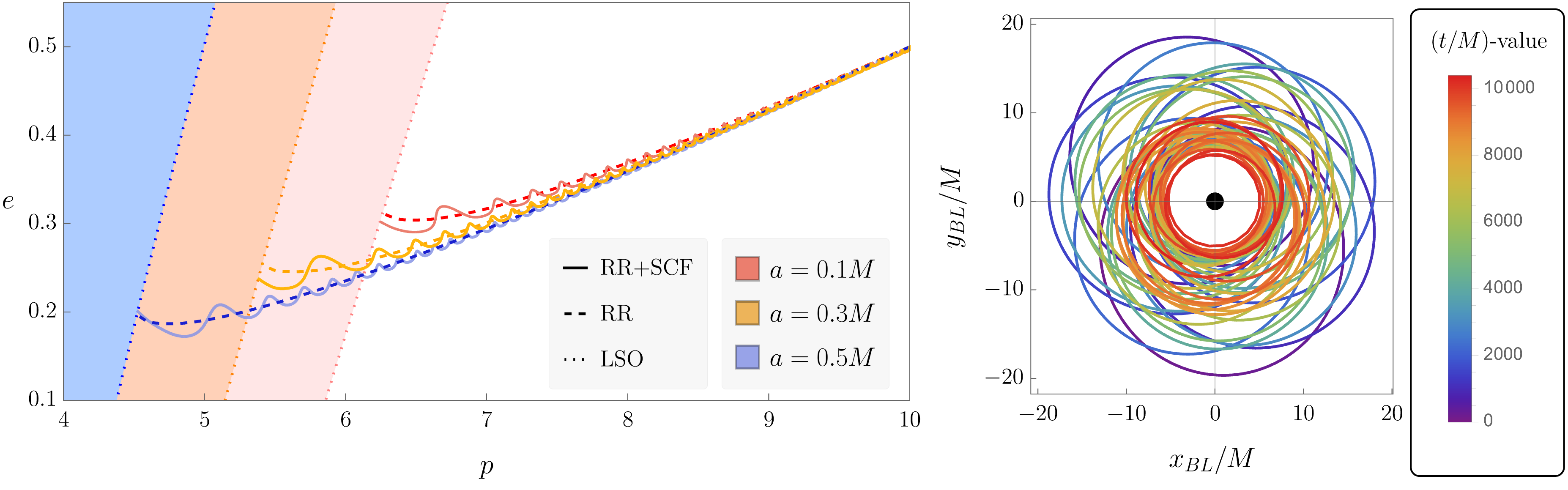}}
    \caption{Equatorial inspiral of a small body with spin aligned with that of the larger black hole.  Left panel shows $p$ versus $e$ for $a=0.1M$ (red), $a=0.3M$ (orange) and $a=0.5M$ (blue).  Solid curves show inspiral in the $(p,e)$-plane for a spinning body; dashed curves show inspiral of non-spinning bodies.  The dotted curves show the last stable orbit for these three black hole spins.  The right panel shows the spinning-body $a = 0.1M$ worldline in the $(x_{BL}, y_{BL})$-plane; color scale encodes the trajectory's evolution (early times in purple, late in red).  In all cases, the inspirals have initial eccentricity $e = 0.5$ and initial semilatus rectum $p = 10$.  We use mass-ratio $\varepsilon = 10^{-2}$ and $s = 1$. }
    \label{fig:avalues}
\end{figure*}

\begin{figure}
\centerline{\includegraphics[scale=0.18]{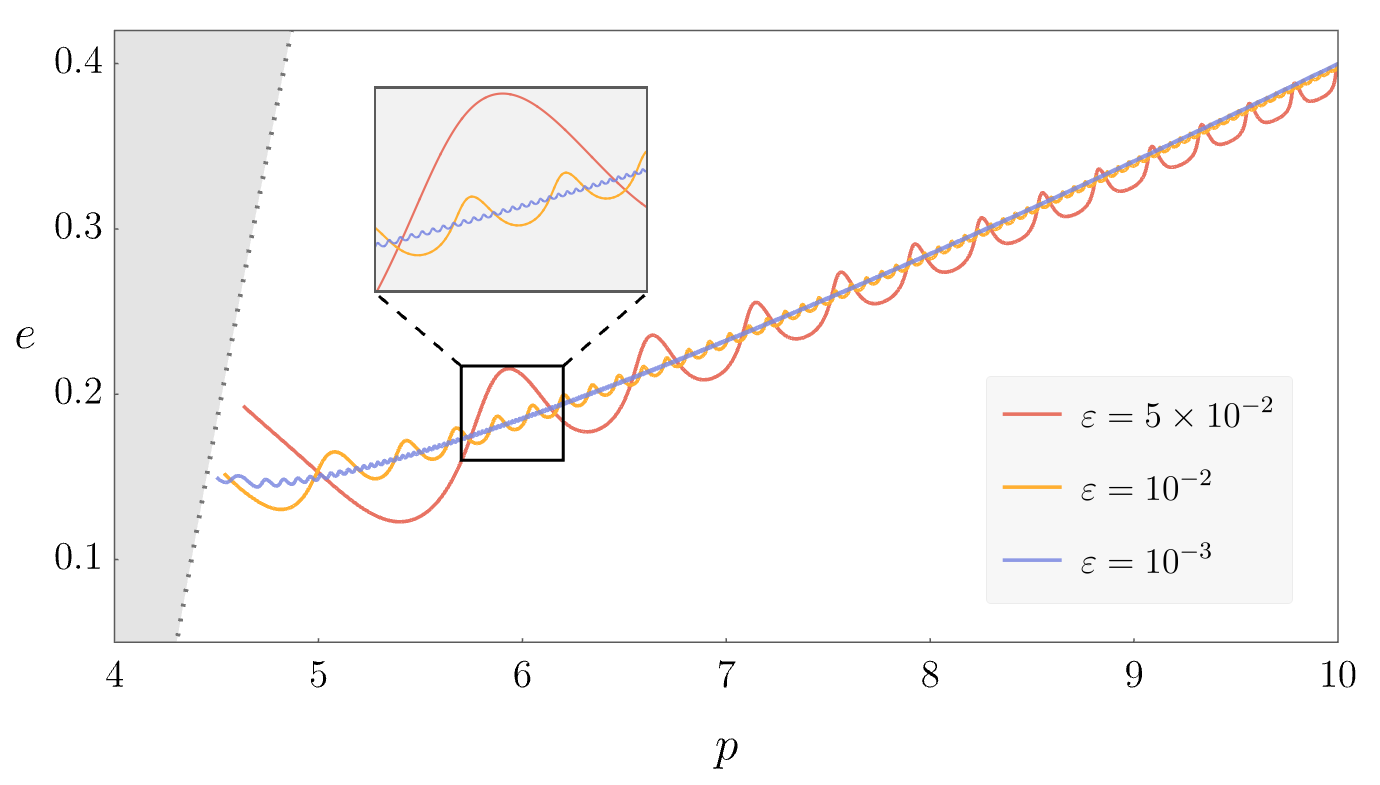}}
    \caption{Evolution of $p$ versus $e$ of an equatorial inspiral for different mass ratios: $\varepsilon = 5 \times 10^{-2}$ (red), $10^{-2}$ (orange), and $10^{-3}$ (blue).  The inset shows a close-up of the region around $p = 6$. The LSO is shown by the gray dotted line.  In all cases, the black hole spin is $a = 0.5M$, initial eccentricity is $e = 0.4$, initial semilatus rectum $p = 10$, and the small-body spin is $s = 1$.}
    \label{fig:massratioinspiraleq}
\end{figure}

\begin{figure}
\hspace{-1cm} 
\centerline{\includegraphics[scale=0.49]{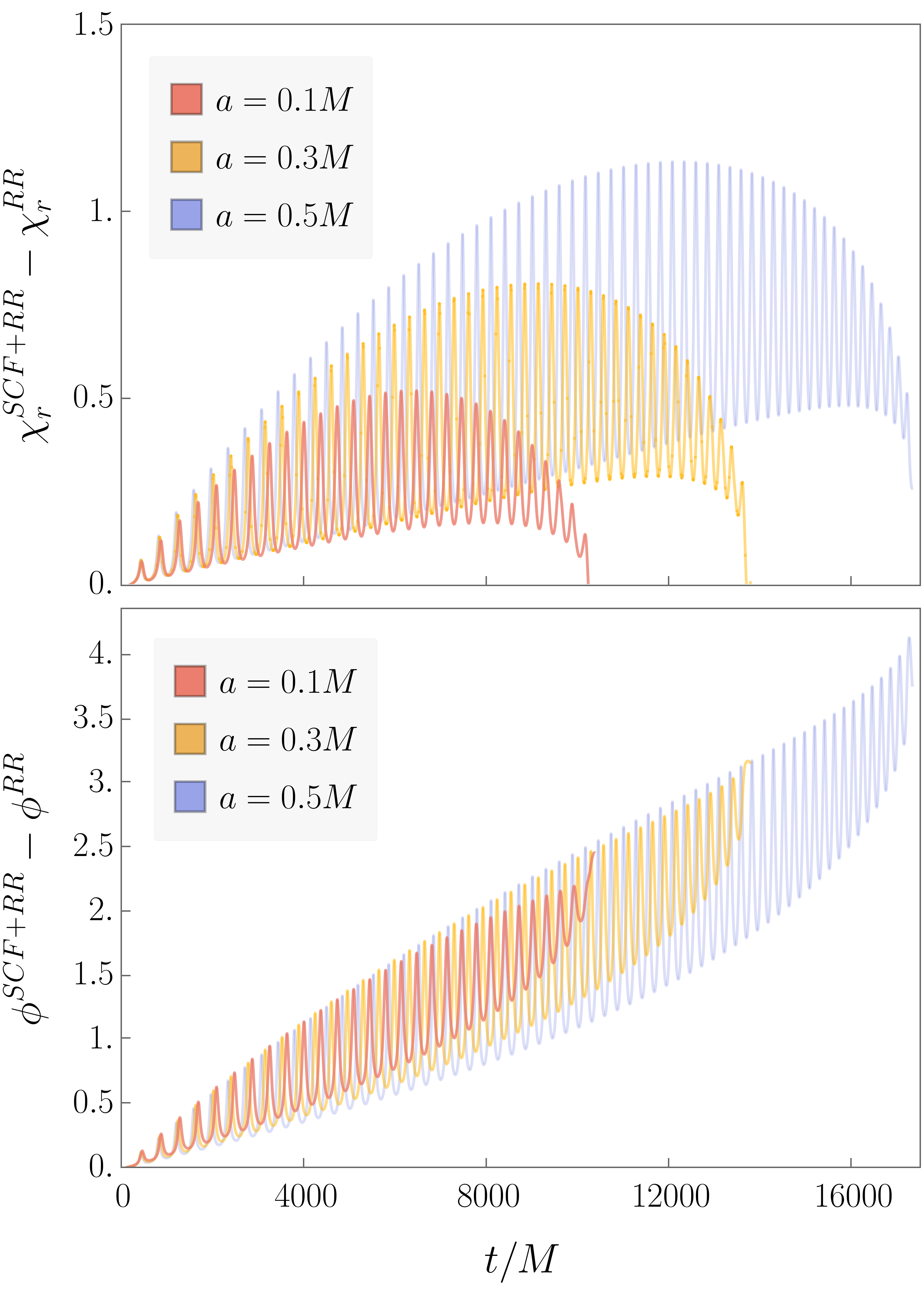}}
    \caption{Dephasing of radial phase $\chi_r^{SCF+RR}-\chi_r^{RR}$ (top panel) and axial phase $\phi^{SCF+RR}-\phi^{RR}$ (bottom panel) for $a = 0.1M$ (red), $a = 0.3M$ (orange) and $a = 0.5M$ (blue). In all cases, initial eccentricity $e = 0.5$, initial semilatus rectum $p = 10$, mass-ratio is $\varepsilon=10^{-2}$, and the small body's spin magnitude is $s = 1$.}
    \label{fig:avaluesphase}
\end{figure}

\subsection{Equatorial inspiral examples}
\label{sec:eqinspirals}

We begin by examining a set of equatorial inspirals with aligned secondary spin.  Each example we consider begins at $p = 10$, $e = 0.5$, and has mass ratio $\varepsilon = 10^{-2}$.  (As mentioned in the introduction we expect astrophysical EMRI systems to have mass ratios of $10^{-4}$ or smaller; we use a larger mass ratio here in order to clearly show spinning body effects.)  We look at inspiral into black holes with $a/M = 0.1$, $0.3$, and $0.5$.  The left-hand panel of Fig.\ \ref{fig:avalues} shows these inspirals in the $(p, e)$ plane.  In all cases, $p$ decreases due to radiation reaction until the system reaches the LSO (showed as a dotted line in this plane); $e$ decreases for much of the inspiral, showing an uptick near the LSO (a well-known strong-field characteristic of GW driven inspiral \cite{Loutrel2019}).  The dashed lines show purely radiation-driven inspirals; the solid lines also include the spin-curvature force for this aligned configuration (computed for a maximally spinning small body: $s = 1$).  The right-hand panel of Fig.\ \ref{fig:avalues} shows the spinning inspiral in the equatorial plane (defining $x_{BL} = r\cos\phi$, $y_{BL} = r\sin\phi$, where $r$ and $\phi$ are the Boyer-Lindquist coordinates of the body's inspiral).  Notice that the late-time geometry of the orbit is nearly circular, consistent with the significant reduction in eccentricity over the inspiral.


In Fig.\ \ref{fig:avalues}, we plot only the evolution of principle orbital elements $p$ and $e$, not including information about the positional orbital element $\chi_r$.  Our expectation is that their secular evolution will only be impacted by the dissipative radiation-reaction force.  This is indeed what we observe: When we include the conservative spin-curvature force, we see oscillations about the radiation-reaction only trajectories (due to the combination of the spin-curvature force with the osculating orbit framework), but the secular evolution remains the same.  Because we are considering a trajectory which is confined to the equatorial plane and with the spin of the small body aligned with the orbit, only harmonics of the radial frequency $\Omega_r$ will be present in the motion.  The oscillations we see in the trajectories shown in Fig.\ \ref{fig:avalues} indeed are periodic with the radial period.

\begin{figure*}
\centerline{\includegraphics[scale=0.45]{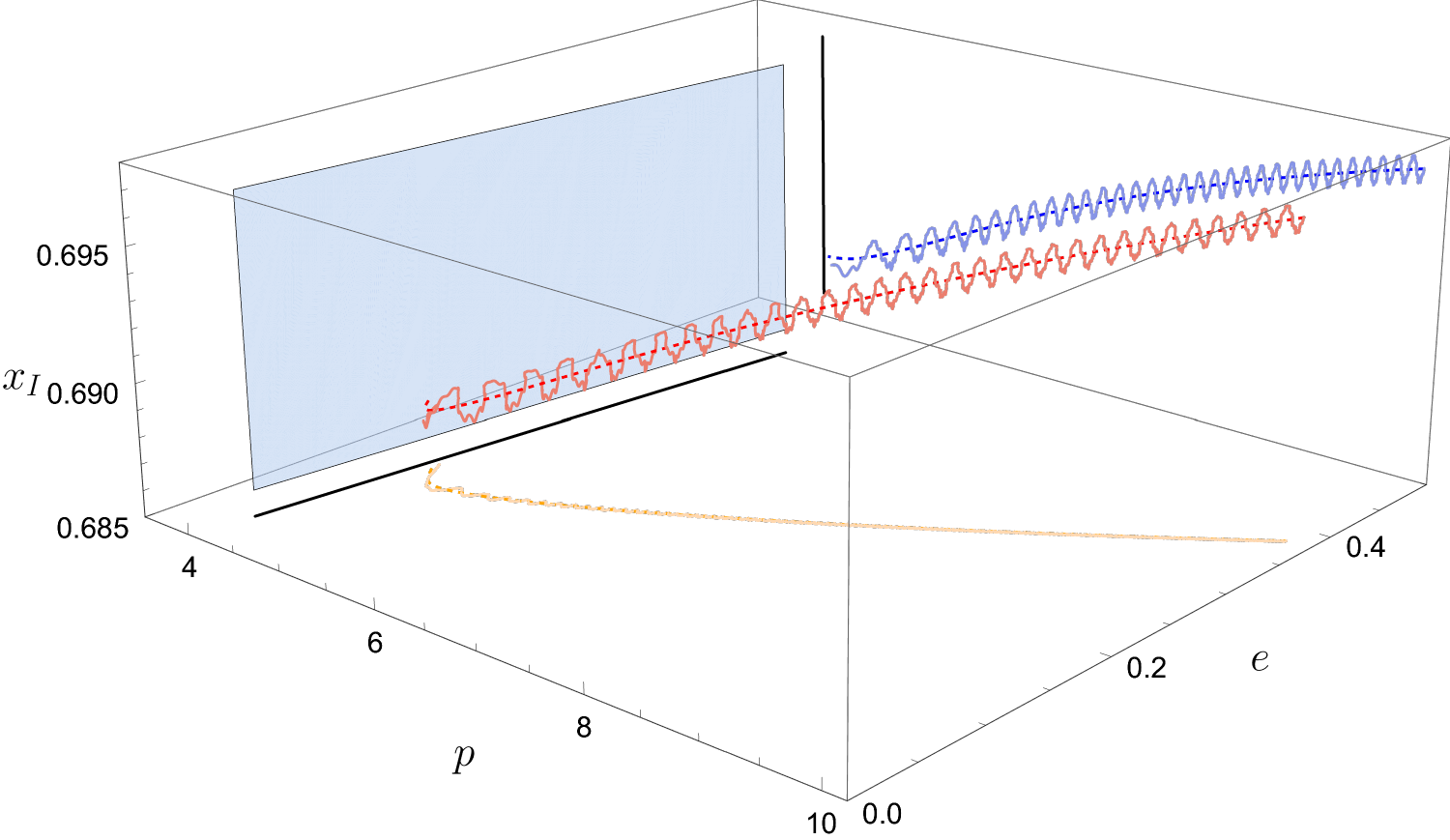}}
    \caption{The trajectory in $p$-$e$-$x_I$ space for an example generic inspiral. This inspiral (red curve) begins at $(p, e, x_I) = (10, 0.38, 0.6967)$ and ends at the LSO (the light blue plane).  The dashed curves show a non-spinning body's inspiral; solid curves are for the inspiral of a spinning small body.  The orange curves show the projection of the inspiral onto the $p$-$e$ plane; the solid black line in this plane is the projection of the last stable orbit at the final value of $x_I$.  (This projection is the same as the top panel of Fig.\ \ref{fig:2Dinspiral}.)  The blue curves show the projection of the inspiral onto the $p$-$x_I$ plane; the solid black curve in this plane is the projection of the LSO at the final value of $e$.  (This projection is the same as the middle panel of Fig.\ \ref{fig:2Dinspiral}.)  We use mass-ratio $\varepsilon = 0.005$ and small-body spin $s = 1$, with $s_\parallel = 0.9$ and $\phi_s = \pi/2$.  See Figure 15 in Ref.\ \cite{Hughes2021} for comparison.}
    \label{fig:3Dinspiral}
\end{figure*}

 Figure \ref{fig:massratioinspiraleq} shows that the amplitude of the oscillations scales with the the mass ratio $\varepsilon$.  For mass ratios corresponding to EMRIs ($\varepsilon \sim 10^{-7}\text{--}10^{-4}$), the oscillations cannot be discerned when plotted; we use values of $\varepsilon$ (e.g., $10^{-2}$ and $5\times10^{-3}$) large enough to make the oscillations visible.  Notice also that the number of oscillations increases inversely with mass ratio.  This is because the duration of inspiral scales inversely with $\varepsilon$, changing the number of orbital cycles the inspiral passes through before reaching the LSO.

Conservative forces affect the evolution of the phases, i.e., the positional orbital elements. In Fig.\ \ref{fig:avaluesphase}, we show the difference in radial phase $\chi_r^{SCF+RR} - \chi_r^{RR}$ (top panel), as well as the difference in axial angle $\phi_r^{SCF+RR} - \phi_r^{RR}$ (bottom panel), accumulated by a spinning body's inspiral and by a non-spinning body's inspiral.  Both of these quantities show both an overall secular trend, and short period oscillations.  The secular trend we find in the radial phase is not monotonic: for $a/M = 0.1$, the dephasing increases to a maximum of about 0.3 radians before dropping back to zero.  Similar trends are seen in the data for $a/M = 0.3$ and $a/M = 0.5$, with the maximum dephasing increasing to about 0.5 radians and 1 radian respectively.  By contrast, the secular dephasing in the axial angle $\phi$ that we find evolves monotonically. 

In addition to these long-term secular trends, which evolve on a radiation reaction timescale $\sim M/\varepsilon$, each dephasing exhibits oscillations on shorter time scales $\sim M$.  For the large-mass ratio systems we are interested in, we are generally only concerned with the longer timescale secular evolution.  A useful way to isolate the secular evolution is via the near-identity transformation (NIT) \cite{Lynch2022,Lynch2021,vandeMeent2018_2}.  In a companion article, we apply this technique to spinning-body inspirals \cite{Drummondinprep2}.  The orbital frequencies $\Omega_r(t)$, $\Omega_\theta(t)$ and $\Omega_\phi(t)$, which evolve during the inspiral, can then be calculated from the NIT equations of motion.  These can then be used as part of the input for generating multivoice inspiral waveforms \cite{Hughes2021}, which we also compute in our companion article \cite{Drummondinprep2}. 

\subsection{Generic inspiral example}
\label{sec:geninspiral}

\begin{figure*}
\centerline{\includegraphics[scale=0.47]{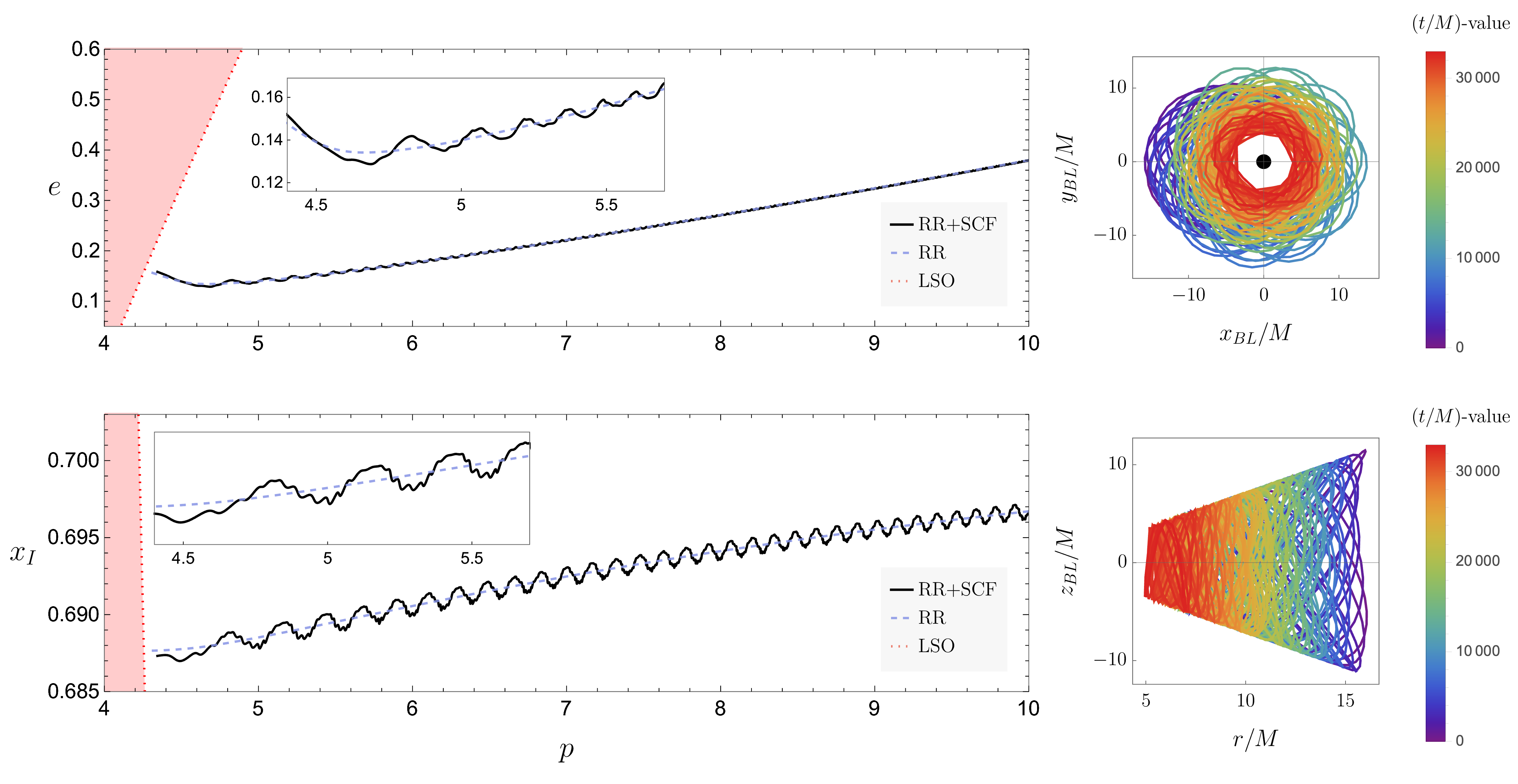}}
    \caption{Evolution of $p$ versus $e$ (top left) and evolution of $p$ versus $x_I$ (middle left) for the inspiral shown in Fig.\ \ref{fig:3Dinspiral}.  Solid black curves show spinning body inspiral; blue dashed curves show non-spinning body inspiral.  In both plots, the last stable orbit (LSO) is shown by the red dotted curve.  The insets show close-ups of inspiral near the LSO.  Top right and middle right panels show projections of the worldline onto the $x_{BL}$-$y_{BL}$ and $r$-$z_{BL}$ planes, with color encoding the time evolution (early times in purple and late times in red). Parameters are identical to those used in Fig.\ \ref{fig:3Dinspiral}.}
    \label{fig:2Dinspiral}
\end{figure*}

\begin{figure}
\centerline{\includegraphics[scale=0.18]{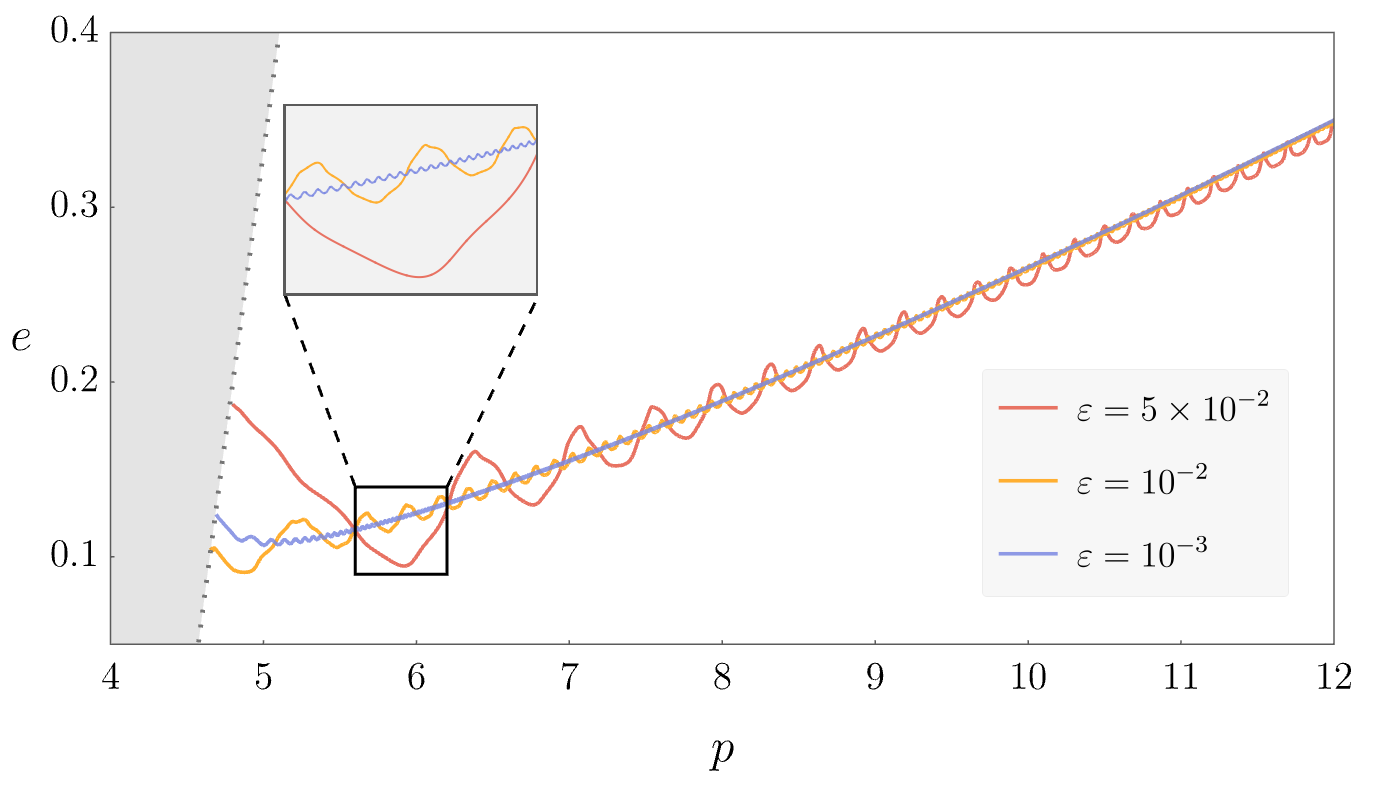}}
    \caption{Evolution of $p$ versus $e$ of a generic inspiral for different mass ratios: $\varepsilon = 5 \times 10^{-2}$ (red), $10^{-2}$ (orange), and $10^{-3}$ (blue).  The inset shows a close-up of the region around $p = 6$. The LSO is shown by the gray dotted line. For all cases, we use black hole spin $a = 0.7M$, initial eccentricity $e = 0.35$, initial inclination $x_I = 0.5$, and initial semilatus rectum $p = 12$. The small-body spin has $s = 1$ and $s_\parallel = s$.   }
    \label{fig:massratioinspiral}
\end{figure}

\begin{figure}
\hspace{-1cm} 
\centerline{\includegraphics[scale=0.46]{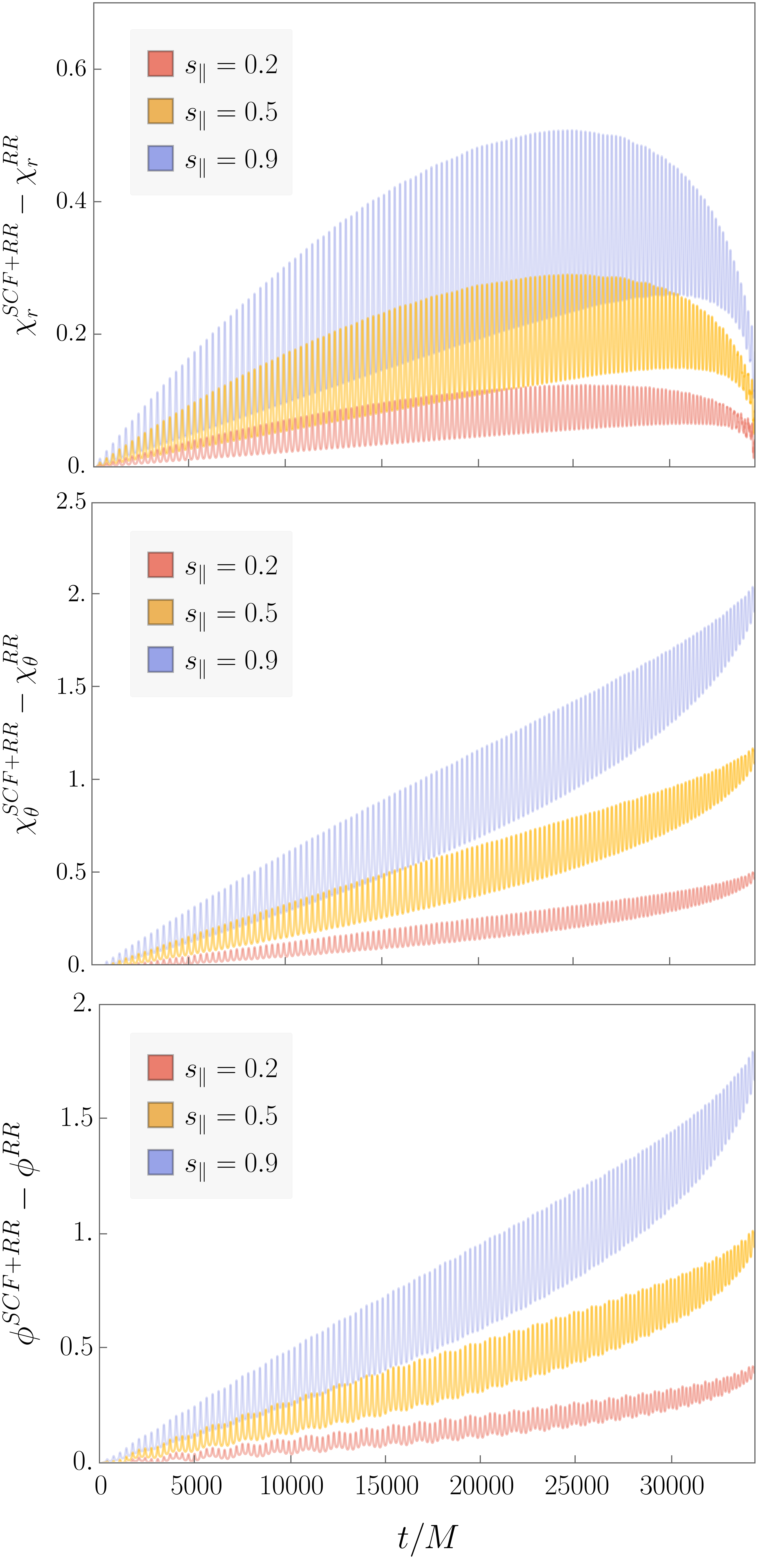}}
    \caption{Dephasing of radial phase $\chi_r^{SCF+RR} - \chi_r^{RR}$ (top panel), polar phase $\chi_\theta^{SCF+RR} - \chi_\theta^{RR}$ (middle panel) and axial phase $\phi^{SCF+RR} - \phi^{RR}$ (bottom panel) for $s_\parallel = 0.9$ (blue), $s_\parallel = 0.5$ (orange) and $s_\parallel = 0.2$ (red).  In all cases, we put black hole spin $a = 0.7M$, initial eccentricity $e = 0.38$, initial inclination $x_I = 0.6967$, initial semilatus rectum $p = 10$, and mass-ratio $\varepsilon = 0.005$.  The small-body spin is characterized by $s = 1$ and $\phi_s=\pi/2$.   }
    \label{fig:avaluesphasegeneric}
\end{figure}

We next look at an example of a generic --- inclined, eccentric, and arbitrarily oriented --- spinning body inspiral.  The red curves in Fig.\ \ref{fig:3Dinspiral} show a generic inspiral, both with (solid line) and without (dashed line) the spin-curvature force.  The orange curve shows the projection of the inspiral onto the $p$-$e$ plane; the blue curve shows the projection onto the $p$-$x_I$ plane. As in Fig.\ \ref{fig:avalues}, the projection onto the $p$-$e$ plane shows a decrease in eccentricity throughout most of the inspiral and then ticks up shortly before reaching the LSO (depicted by a black line).  The inspiral increases in inclination (corresponding to a decrease in $x_I$) all the way to the LSO, with no deep strong-field reversal of sign unlike the $p$-$e$ trajectory.

Figure \ref{fig:2Dinspiral} shows a more detailed depiction of the projections of the inspiral onto the $p$-$e$ and $p$-$x_I$ planes (leftmost panels of the first two rows).  Each of these panels includes an inset which zooms in on the inspiral close to the LSO. Just as for Fig.\ \ref{fig:avalues}, the secular evolution of the principle orbital elements $p$, $e$ and $x_I$ is unaffected by the presence of the spin-curvature force, but inclusion of the spin-curvature force leads to oscillations about the secular trajectory.  Notice that the generic inspiral has harmonic structure at multiple timescales --- the oscillations have a rather more complicated structure than we saw in the case of aligned equatorial inspirals.  This more intricate harmonic structure is because there are terms in the equations of motion which are periodic with the four frequencies $\Omega_r$, $\Omega_\theta$, $\Omega_\phi$ and $\Omega_s$. Harmonics at frequency $\Omega_s$ are due to the precession of the small-body's spin vector; these harmonics can be seen in the components of the small body's spin which are shown in the bottom two rows of Fig.\ \ref{fig:2Dinspiral}.  The most clear oscillation in the $p$-$e$ trajectory is at the radial frequency (which is initially $M\Omega_r = 0.0186359$); the oscillations in the $x_I$-$p$ trajectory are more complex, involving beats between the four frequencies.
 
The top two rows of the rightmost column of Figure \ref{fig:2Dinspiral} display the evolution of the trajectory projected onto the $x_{BL}$-$y_{BL}$ plane (first row) and $r$-$z_{BL}$ plane (second row); we have defined $x_{BL} = r\sin\theta\cos\phi$, $y_{BL} = r\sin\theta\sin\phi$, $z_{BL} = r\cos\theta$, with $r$, $\theta$, and $\phi$ the Boyer-Lindquist coordinates along the inspiral. The $x_{BL}$-$y_{BL}$ inspiral projection shares similar features to the corresponding equatorial version in Fig.\ \ref{fig:avalues}.  In the $r$-$z_{BL}$ inspiral projection, we see that the maximum $|z_{BL}|$ decreases as inspiral progresses.  Although the inclination angle $I$ increases during inspiral, the effect is quite small.  The shrinking of $r$ due to radiative backreaction is much more significant, so $|z_{BL}| = |r\cos\theta|$ decreases overall.  In Fig.\ \ref{fig:massratioinspiral}, we show how the amplitude of the oscillations changes with mass ratio. As in Figure \ref{fig:massratioinspiraleq}, the amplitude of oscillations decreases with decreasing $\varepsilon$, and the number of oscillations increases.  This is because smaller mass ratio systems take longer to inspiral, undergoing more orbital cycles before reaching the LSO.
 
Similar to the equatorial case shown in Fig.\ \ref{fig:avaluesphase}, the effect of the spin-curvature force is apparent in the net change to the phase evolution. Fig.\ \ref{fig:avaluesphasegeneric} displays the difference in radial phase $\chi_r^{SCF+RR}-\chi_r^{RR}$, polar phase $\chi_\theta^{SCF+RR}-\chi_\theta^{RR}$ and axial angle $\phi_r^{SCF+RR}-\phi_r^{RR}$ between a spinning and non-spinning small body throughout the inspiral. Again, we see that the secular evolution of the spin-curvature-induced dephasing of $\chi_r$ is not monotonic, while the dephasing of $\chi_\theta$ and $\phi$ is monotonic on average. We also show how the dephasing depends on the proportion of the small body's spin that is aligned with the orbital angular momentum $s_\parallel$. We see that the amount of dephasing is linearly related to $s_\parallel$; as $s_\parallel$ increases, the size of the dephasing increases by the same factor. For example, when $s_\parallel=0.9$ (blue curve), $\chi_\theta^{SCF+RR}-\chi_\theta^{RR}$ reaches a value of around 2 by the end of the inspiral, while for $s_\parallel=0.5$ (blue curve), $\chi_\theta^{SCF+RR}-\chi_\theta^{RR}$ reaches a value of a little more than 1.

\section{Conclusions and future work}
\label{sec:conclusions}

In this analysis, we use an osculating geodesic framework to compute completely generic inspirals of spinning bodies for the first time.  Our analysis has multiple potential applications. Precise models of large mass-ratio black hole binaries need to be developed for measurements by the future space-based GW detector LISA. In addition, studies of the effect of small-body spin could be a useful foundation for building surrogate models of GWs from black hole binaries with less extreme mass ratios.  In a companion article, we apply a near-identity transformation to isolate the secular effects which allows for very fast computation and calculation of the orbital phases, which we use as input for generating multi-voice gravitational waveforms \cite{Drummondinprep2}.

In our calculations, we omit key small-body effects which are known will be of observational importance, including the conservative and oscillating dissipative first-order self-force, the averaged dissipative second-order self-force, and the back-reaction due to the GW flux associated with the spin of the small body.  In the inspirals we present here, we compute the (non-spinning) point-particle GW fluxes using the Teukolsky equation \cite{Hughes2021} and then add spin-curvature forces using an osculating geodesic framework in order to construct a spinning-body worldline.  We thus exclude the impact of the dipole component of the stress-energy tensor.

A more complete framework for spinning-body inspirals will include contributions to the GW fluxes due to small-body spin. Such a scheme was implemented for equatorial worldlines by Skoup\'{y}  and Lukes-Gerakopoulos in Refs.\ \cite{Skoupy2021} and \cite{Skoupy2022}. In these works, the authors compute the spinning-body GW fluxes \cite{Skoupy2021} and build the corresponding equatorial inspiral \cite{Skoupy2022}.  They achieve this by using an osculating \textit{spinning-body orbit} framework, i.e., the elements parameterizing the osculating orbits are those of spinning-body orbits rather than bound geodesics. Skoup\'{y}  and Lukes-Gerakopoulos have recently computed the asymptotic GW fluxes from a spinning body on generic orbits \cite{Skoupy2023}. Extending this analysis to generate an inspiral requires efficient computation of the orbital frequencies of generic spinning-body orbits; the approach used in Ref.\ \cite{Drummond2022_1,Drummond2022_1} can be computationally slow.  Developing an efficient scheme for computing spinning-body frequencies is likely to be useful for generic inspiral calculations. One approach involves using Chebyshev polynomial interpolation across parameter space.

Once this calculation is extended to fully generic inspirals, we plan to conduct a detailed and systematic comparison with the calculation in this article to establish how large of an effect the inclusion of small-body spin in the flux calculation has on the trajectory. In general, comparison of the significance of the different post-adiabatic effects across parameter space would allow us to identify if there are any approximations we can make which have a minimal effect on the accuracy of the calculation but lead to a substantial reduction in computational expense. 

Along these lines, one approximation we plan to investigate involves computing the Teukolsky fluxes for a point-particle trajectory and shifting the frequencies using the spin-curvature corrections described in Ref.\ \cite{Drummond2022_1,Drummond2022_1} rather than recomputing the entire spinning-body trajectory for each point in parameter space. This would neglect the oscillatory correction to the true anomaly angles associated with the spinning-body orbit.  We suspect that errors due to this neglect will be small enough that this could make a useful approximation for computing waveforms from spinning body orbits.  However, a thorough study needs to be conducted to assess this.

\section*{Acknowledgements}

LVD, DRB and SAH were supported by NASA ATP Grant 80NSSC18K1091 and NSF Grant PHY-2110384; AGH was supported by ATP Grant 80NSSC18K1091 while at MIT.  This work makes use of the Black Hole Perturbation Toolkit.  LVD would like to thank Philip Lynch for generously sharing his code, which was used on test cases for comparisons with our own code.

\bibliographystyle{unsrt}

\bibliography{SpinningBodyInspiral}
\end{document}